\documentclass[twocolumn]{aastex631}

\newcommand\msunyr{$M$\mbox{$_{\normalsize\odot}$}~\rm{yr}$^{-1}$}
\newcommand\msun{$M$\mbox{$_{\normalsize\odot}$}}

\newcommand\lbol{$L$\mbox{$_{\rm bol}$}}

\newcommand\teff{$T_{\rm eff}$}

\newcommand\logl{$\log (L/L_\odot)$}

\newcommand\N[1]{{\color{blue} \bf #1}} 

\usepackage{soul}

\begin{document}

\title{JWST reveals a luminous infrared source at the position of the failed supernova candidate N6946-BH1}

\author[0000-0003-4666-4606]{Emma R.\ Beasor}\altaffiliation{Bok Fellow}
\affil{Steward Observatory, University of Arizona, 933 North Cherry Avenue, Tucson, AZ 85721-0065, USA}

\author[0000-0002-0832-2974]{Griffin Hosseinzadeh}
\affil{Steward Observatory, University of Arizona, 933 North Cherry Avenue, Tucson, AZ 85721-0065, USA}

\author[0000-0001-5510-2424]{Nathan Smith}
\affil{Steward Observatory, University of Arizona, 933 North Cherry Avenue, Tucson, AZ 85721-0065, USA}

\author[0000-0002-2010-2122]{Ben Davies}
\affil{Astrophysics Research Institute, Liverpool John Moores University, Liverpool Science Park ic2, 146 Brownlow Hill, Liverpool L3 5RF, UK}

\author[0000-0001-5754-4007]{Jacob E.\ Jencson}
\affil{Department of Physics and Astronomy, Johns Hopkins University, 3400 North Charles Street, Baltimore, MD 21218, USA}
\affil{Space Telescope Science Institute, 3700 San Martin Drive, Baltimore, MD 21218, USA}

\author[0000-0002-0744-0047]{Jeniveve Pearson}
\affil{Steward Observatory, University of Arizona, 933 North Cherry Avenue, Tucson, AZ 85721-0065, USA}

\author[0000-0003-4102-380X]{David J.\ Sand}
\affil{Steward Observatory, University of Arizona, 933 North Cherry Avenue, Tucson, AZ 85721-0065, USA}

\begin{abstract}

N6946-BH1 is the first plausible candidate for a failed supernova (SN), a peculiar event in which a massive star disappears without the expected bright SN, accompanied by collapse into a black hole (BH). Following a luminous outburst in 2009, the source experienced a significant decline in optical brightness, while maintaining a persistent infrared (IR) presence. While it was proposed to be a potential failed SN, such behaviour has been observed in SN impostor events in nearby galaxies. Here, we present late-time observations of BH1, taken 14 years after disappearance, using JWST’s NIRCam and MIRI instruments to probe a never-before-observed region of the object’s spectral energy distribution. We show for the first time that all previous observations of BH1 (pre- and post-disappearance) are actually a blend of at least 3 sources. In the near-IR, BH1 is notably fainter than the progenitor but retains similar brightness to its state in 2017. In the mid-IR the flux appears to have brightened compared to the inferred fluxes from the best-fitting progenitor model. The total luminosity of the source is between 13 -- 25\% that of the progenitor. We also show that the IR SED appears consistent with PAH features that arise when dust is illuminated by near-ultraviolet radiation.  At present, the interpretation of N6946-BH1 remains uncertain. The observations match expectations for a stellar merger, but theoretical ambiguity in the failed SN hypothesis makes it hard to dismiss.

\end{abstract}


\section{Introduction} \label{sec:intro}
Pre-explosion imaging has provided a unique opportunity to directly link progenitors to their stellar explosions, tightly constraining stellar evolutionary models. For single stars, theory predicts that all stars with initial masses between $8 - 25\,M_{\odot}$ will end their lives in the red supergiant (RSG) phase, ultimately exploding as H-rich Type II-P SNe. RSGs have been observed as the progenitors to Type II-P SNe many times \citep[e.g.][]{smartt2015observational, vandyk2017direct,jencson2023ixf}, but so far no progenitors have been observed at the upper end of the mass range ($\sim 19-25\,M_{\odot}$). This `mass gap' of detected SN progenitors has been termed `The Red Supergiant Problem'. There is still debate in the literature over the statistical significance of this result \citep[e.g.][]{davies2020red,davies2020on, kochanek2020on}, and the fate of RSG stars with intitial masses above $19\,M_{\odot}$ remains a mystery.

A number of groups have attempted to reconcile the RSG problem. It has been suggested that enhanced mass-loss could drive the high mass RSGs back to the blue of the HRD, forcing them to explode as different classes of SNe \citep[e.g.][]{georgy2012yellow}, or that these objects enshroud themselves in a thick layer of dust, obscuring their light and causing the stars to appear less luminous \citep[and hence, inferred erroneously to be less massive e.g.][]{beasor2016evolution}. {If those more massive RSGs explode while enshrouded, they may appear as SNe IIn instead of SNe II-P \citep{smith2009red}, and there are plenty of SNe IIn by number to make up for the missing SNe~II-P \citep{smith11}. Dust-enshrouded RSGs are, however, very rare, and this phase would need to be extremely short-lived \citep{beasor2022dersgs}.

Perhaps the most exotic solution is the possibility that the highest mass RSGs collapse directly to a black hole (BH) with little or no visible explosion \citep{smartt2009death}. This idea has some theoretical support, with a number of papers claiming that some RSGs may be more difficult to explode due to their core structure \cite[e.g.][]{lovegrove2013very, sukhbold2018high}. The mass range of RSGs that seem to be missing from IIP progenitor populations roughly matches the mass range that was subsequently predicted to suffer implosion due to core compactness \citep[see e.g.][]{sukhbold2018high}. If confirmed, the collapse of RSGs directly to black holes may also provide another avenue for the production of stellar-mass BHs, such as those observed via the LIGO/Virgo interferometers \citep[e.g.][]{abbott2017gw1}, of which there are now more than 100 gravitational wave observations of stellar-mass BHs with masses $< 30\,M_{\odot}$. This would fundamentally alter our view of stellar evolutionary theory, and open new avenues for the formation of stellar-mass BHs. While stellar-mass BHs are clearly being formed, the actual range of stellar initial masses that lead to these BHs requires observational constraints. Such constraints remain ambiguous and highly debated.

While a BH-formation mass threshold of $\sim 20\,M_{\odot}$ would potentially explain both the RSG problem and the LIGO/VIRGO observations of stellar-mass BHs \citep{kochanek2015constraints}, there has been no direct observation of a star collapsing to a BH. The possibilities of directly observing these RSGs collapse are limited, since their rate is expected to be low \cite[$\sim 1$/year within 30 Mpc,][]{davies2020red}. If the BH mass threshold {\it is} 20 \msun, and everything above this mass fails to explode, we would expect the failed SN rate to be more than 40\% of the total successful SNe rate \citep{smith11}. If it is only RSGs in the 20--30\msun\ range, the expected rate falls to $\sim$ 20\% of the CCSNe rate. One way to detect such an event is to continually monitor nearby galaxies and search for stars that disappear with little or no visible transient \citep{kochanek2008survey}. Given the low rates of these events, the confirmation of even one disappearing RSG would be a significant discovery. 

The first plausible failed SN candidate has now been identified in the nearby star-forming galaxy NGC~6946, which has hosted 10 observed CCSNe in the last century. This source has been presumptively named N6946-BH1 \citep{gerke2015search, adams2017search,neustadt2021lbt,basinger2021bh1}. Pre-explosion imaging revealed a candidate progenitor that was interpreted as a $20-25\,M_{\odot}$ RSG ($\log(L/L_{\odot}) = 5.3-5.5$), near the top end of the RSG mass range, and a mass which is yet to be directly observed as a progenitor to a SN. The star was monitored for many epochs from 2007, showing mild but normal RSG variability. In 2009, the star experienced a moderately luminous optical outburst that had a peak luminosity about 10x higher than the quiescent RSG. Following this, the optical brightness declined steeply to below the progenitor brightness. As of 2015, the star appeared 5 magnitudes fainter in the optical, while appearing to brighten in the mid-IR ($\lambda$ $>$ 5$\mu$m). When comparing the pre- and post-disappearance HST photometry (F606W \& F814W), it is clear the progenitor is far fainter in the late-time imaging, with the star essentially disappeared in the F606W (0.59~$\mu$m) filter, while there is some weak emission visible in the F814W (0.79~$\mu$m) filter. In addition, there is clear emission in the F110W and F160W HST filters.

\citet{adams2017search} discussed two competing explanations for the apparent disappearance. The first is that N6946-BH1 has in fact collapsed to a BH and is no longer present at any wavelength. If true, this would be the first directly observed case of an RSG collapsing to BH with no visible explosion. Their second hypothesis is that the RSG progenitor has experienced an extreme mass-loss event, covering itself in a thick spherical layer of dust and making itself invisible at optical wavelengths, but re-radiating its luminosity at longer mid-IR wavelengths.  Noting that most mechanisms for stellar outbursts in binary systems should be inherently non-spherical, \citep{kashisoker17} proposed a different version of the second hypothesis, where an ejected dust shell has a toroidal geometry and is seen nearly edge-on.  They predicted that a mid-IR source would remain but with a somewhat lower integrated luminosity than the progenitor.

Considerable uncertainty remains about the fate of N6946-BH1. First and foremost, collapse to a BH is not the only reason that a star might fade at visual wavelength. Although the star faded significantly at visual wavelengths, it has shown little change in flux at 3-5~$\mu$m in Spitzer data. Since the star did produce a transient that resembles some non-terminal eruptive variables, it is plausible that this eruptive event ejected mass and formed dust. In that case, the star may have faded at visual wavelengths because of obscuration by a dust shell, not because it collapsed to a BH. The most recent Spitzer and HST images (all taken in 2018 and included in Fig.~\ref{fig:SED}, Basinger et al. 2021) are still unable to differentiate between these two scenarios because they lack access to crucial mid-IR wavelengths. A dust shell should emit brightly at wavelengths $>$5$\mu$m \citep[due to the silicate emission feature seen in RSGs;][]{beasoe2020new}, a region of the SED now only accessible by JWST.

Here we present JWST NIRCam and MIRI observations of N6946-BH1. In Section \ref{sec:observations} we discuss the observations and data reduction. In Section \ref{sec:sed} we discuss the shape of the spectral energy distribution (SED) and estimate a luminosity for the remaining source, as well as exploring various SED fits. In Section \ref{sec:discussion} we discuss our findings in the context of massive stellar evolution. \footnote{While this paper was under review, another paper appeared on the arxiv also focusing on the fate of BH1. For another interpretation of the remaining infrared source, see \citet{kochanek2023bh1}.}

\section{Observations and Data Reduction}\label{sec:observations}
We obtained data from JWST using the Near Infrared Camera \citep[NIRCam, ][]{nircam} and the Mid-Infrared Imager \citep[MIRI,][]{miri2015} on 25 August 2023 for Program 3773 (PI: E. Beasor\footnote{http://dx.doi.org/10.17909/vm7n-vn41}). Table \ref{tab:observations} details the filters and exposure times used. Our observations covered the spectral energy distribution of BH1 from 1.15$\mu$m to 21$\mu$m. For NIRCam we used the {\tt INTRASCA} dither pattern with default offsets and the {\tt FULLP} array (64'' $\times$ 64'') which is optimised for point sources. For MIRI we used the {\tt FULL} array with the {\tt CYCLING LARGE} 4-point dither pattern.

\subsection{Identifying the correct source}\label{sec:source}
To ensure we are looking at the same source as was identified as the progenitor in the pre-disappearance images we use {\tt IDL} packages {\tt GCNTRD} and {\tt starast}\footnote{https://idlastro.gsfc.nasa.gov/} to derive more precise astrometric solutions for each image, including for the pre-disappearance HST F814W image. We begin by identifying common reference stars in each of the images. Since a number of sources that are visible at shorter wavelengths are not visible at longer wavelengths, we use different sets of reference stars for each set of images. Once the reference stars have been chosen, we use {\tt GCNTRD} to locate the precise pixel position of the centroid of each star. This program uses the {\tt DAOPHOT} centroid finding function to locate the centroid by fitting Gaussian functions to the stellar profiles and requires input in the form of a pixel position and a full-width half maximum (FWHM)\footnote{FWHMs for each filter taken from HST and JWST User Documentation. For NIRCam, we used the empirical FWHMs.}. Once the centroids of each reference star have been located, we use {\tt starast} to compute a precise astrometric solution ($\sim$0.001 arcsec) based on the assumption of a gnomonic projection. 

We begin at the shorter wavelengths, using the JWST F115W image as the reference image.  We take the RA and Dec positions of the reference stars in the reference image, and match them to pixel positions of the same reference stars in the current image, in order to compute the astrometric solution and update the header accordingly. This is repeated at each wavelength, using the previous image as the reference image. By the end of process, all of the filters contain the same WCS information.

The most significant difficulty in locating the correct source comes in the shortest NIRCam filters, F115W and F182M. These images reveal 3 sources coincident with the original BH1 progenitor source identified in the pre-disappearance HST WFPC2 F606W and F814W filters (see Fig. \ref{fig:zoom_align}). To determine whether any of these 3 sources are associated with BH1, we use the aligned images and plot the position of the F814W centroid onto the NIRCam data. Figure \ref{fig:zoom_align} shows that the centroid of the F814W source aligns with the southernmost star at the bottom of the triangular group. Indeed, we also find that the centroids of the longer wavelength emission in the NIRCam and MIRI data also align with this star to within 0.02 arcseconds. As such, we assume this object to be the source visible in the pre-disappearance images and that it currently dominates the flux in the infrared, such that we can treat the emission at $\lambda > 2.5 \micron$ as being from a single source.  

Given that these sources were previously unresolved in the pre-disappearance images, we derive fluxes for all 3 objects to assess how much they are likely to have contributed to the inferred luminosity of BH1. We find that the two other stars in the asterism are of a similar brightness to each other, see Table \ref{tab:data}.

\begin{figure}
    \centering
    \includegraphics[width=\columnwidth]{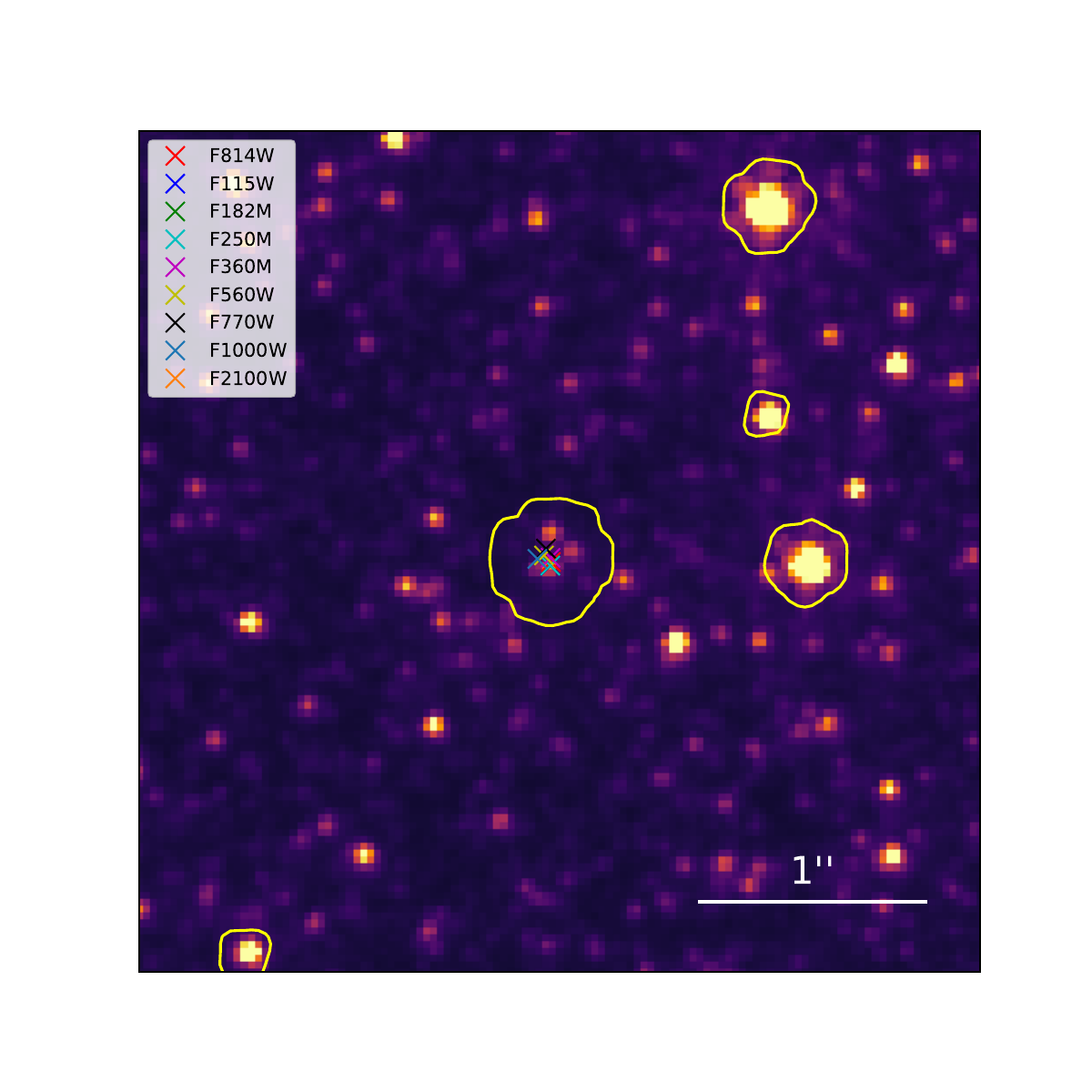}
    \caption{Centroids of the detected source in each filter overplotted onto the NIRCam F115W image. The yellow contours show the source locations in the F814W pre-disappearance data. }
    \label{fig:zoom_align}
\end{figure}

\begin{figure*}
    \centering
    \includegraphics[width=16cm]{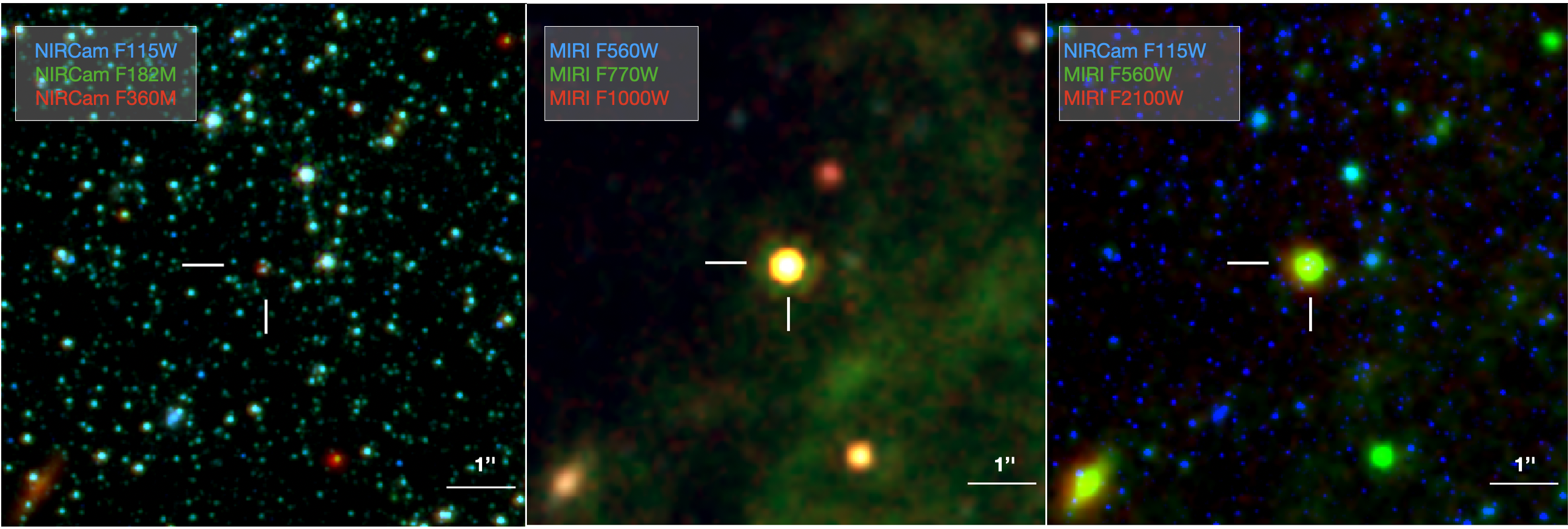}
    \caption{Three colour images of BH1. {\it Left:} NIRCam filters. {\it Middle:} MIRI filters. {\it Right:} NIRCam + MIRI filters. }
    \label{fig:enter-label}
\end{figure*}
\subsection{Extracting the photometry}
For each filter, we downloaded the Level 2 (aligned) and 3 (aligned and stacked) calibrated images (filenames ending with \texttt{i2d.fits}), as well as the Level 3 source catalog (ending with \texttt{cat.ecsv}), from the Mukulski Archive for Space Telescopes and measured photometry. We identified a bright star north by northwest of BH1, at ICRS coordinates $\alpha \approx 308.8641603\degr$, $\delta \approx 60.13608154\degr$, in each source catalog and determined a rigid shift that would match the coordinates of this star in the F115W image, which has the highest resolution. We then applied this shift to each Level 2 exposure, assuming that any pointing discrepancy would be approximately the same from exposure to exposure in the same filter. The star was not present in the source catalogs for F770W or F2100W, so we used the provided coordinate solutions for those exposures as-is.

We performed aperture photometry on BH1 and each of the two nearby sources using Photutils \citep{bradley2016photutils}, following a procedure similar to \cite{hosseinzadeh_jwst_2023}, on each of the Level 2 exposures. For both instruments, we used relatively small apertures with radii that enclose 60\% of the energy in order to avoid contamination from the other sources. We subtracted a background level determined by the median of the pixels in another circular aperture of the same size, offset from the source in a southerly direction with no visible sources by 3.25 radii to avoid contamination from the wings of the PSF. We then applied aperture corrections and zero points for each filter using the JWST Calibration Reference Data System \cite[CRDS, ][]{greenfield2016crds}. Finally, we averaged the background-subtracted fluxes in each of the four Level 2 exposures, weighted by their uncertainties, to obtain a single measurement in each filter, reported in Table~\ref{tab:data}. Fig.~\ref{fig:imaging} shows a cutout around the source in each filter, scaled to the same background-subtracted flux level, with the BH1 and background apertures marked.

\begin{table*}
\centering
\begin{tabular}{lllccc}
\hline
MJD & Instrument & Filter & Exposure time (s) \\
\hline\hline
60181.682 & NIRCam & F115W + F250M & 4208.8 \\
60181.727 & NIRCam & F182M + F360M & 2104.4 \\
60181.757 & MIRI & F560W & 233.1 \\
60181.765 & MIRI & F770W & 122.1 \\
60181.772 & MIRI & F1000W & 122.1 \\
60181.778 & MIRI & F2100W & 122.1 \\
\hline
\end{tabular}
\caption{Observation details for each filter. Note that for NIRCam two images are taken simultaneously. }
\label{tab:observations}
\end{table*}

\begin{table*}
\centering
\begin{tabular}{lllcc}
\hline
Source & Instrument & Filter & Magnitude (AB) & Flux ($\mu$Jy) \\
\hline\hline
BH1 & NIRCam & F115W & $25.288 \pm 0.013$ & $0.279 \pm 0.003$ \\
BH1 & NIRCam & F182M & $24.630 \pm 0.012$ & $0.510 \pm 0.006$ \\
BH1 & NIRCam & F250M & $25.697 \pm 0.020$ & $0.191 \pm 0.004$ \\
BH1 & NIRCam & F360M & $23.603 \pm 0.007$ & $1.314 \pm 0.008$ \\
BH1 & MIRI & F560W & $20.257 \pm 0.005$ & $28.6 \pm 0.1$ \\
BH1 & MIRI & F770W & $19.351 \pm 0.004$ & $66.0 \pm 0.2$ \\
BH1 & MIRI & F1000W & $19.969 \pm 0.008$ & $37.4 \pm 0.3$ \\
BH1 & MIRI & F2100W & $19.162 \pm 0.030$ & $78.6 \pm 2.2$ \\
Source 2 & NIRCam & F115W & $25.427 \pm 0.014$ & $0.245 \pm 0.003$ \\
Source 2 & NIRCam & F182M & $24.793 \pm 0.014$ & $0.440 \pm 0.006$ \\
Source 3 & NIRCam & F115W & $25.832 \pm 0.019$ & $0.169 \pm 0.003$ \\
Source 3 & NIRCam & F182M & $25.152 \pm 0.018$ & $0.316 \pm 0.005$ \\
\hline
\end{tabular}
\caption{Flux and magnitude data for BH1, Source 2, and Source 3 in each filter.}
\label{tab:data}
\end{table*}

\begin{figure*}
    \centering
    \includegraphics[width=\textwidth]{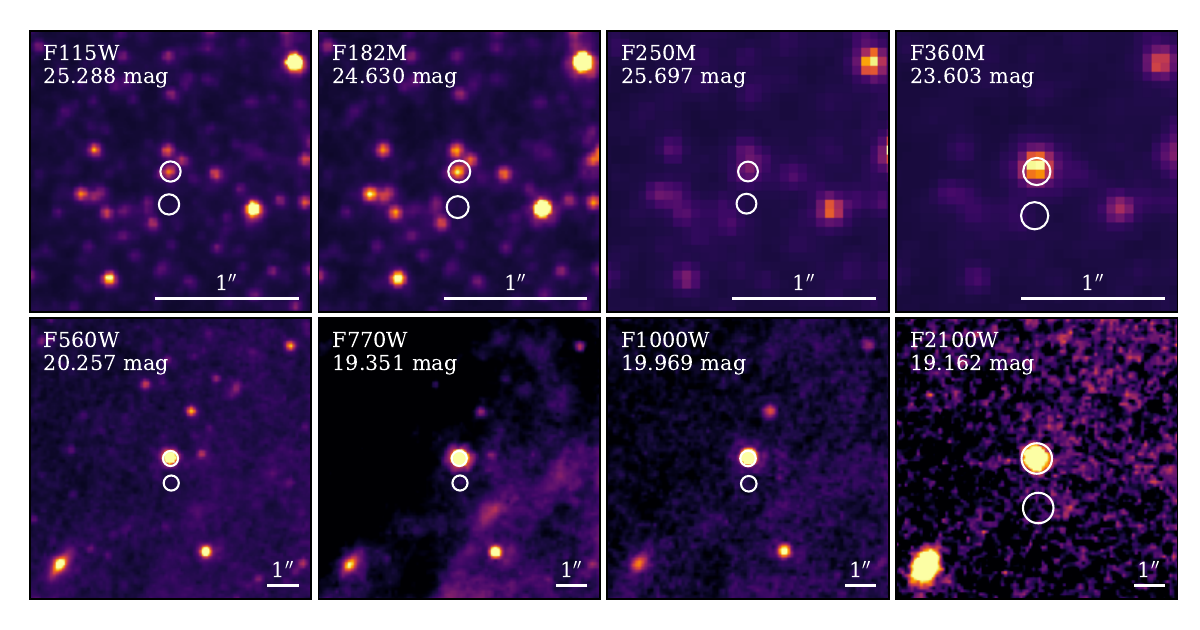}
    \caption{Cutouts of the JWST images centered on BH1. The white circles show the apertures used to determine the fluxes of BH1 and the background. The images have been reprojected so that up is north and left is east and scaled to the same background-subtracted flux level. The filter name and AB magnitudes are shown in each cutout. }
    \label{fig:imaging}
\end{figure*}

\section{The Spectral Energy Distribution}\label{sec:sed}
We have shown that a clear source remains at the position of BH1. In Figure \ref{fig:SED} we show the spectral energy distribution (SED). The JWST data are shown by the filled circles. We also show the 2018 HST and Spitzer data by the open diamonds and squares, respectively. Note the flux in the F115W and F182M filters does not include the contribution from the two other sources identified at short wavelengths. We also show the progenitor photometry taken prior to the 2009 outburst with the red open stars, as well as the best fitting progenitor {\tt DUSTY} model.

\begin{figure}
    \centering
    \includegraphics[width=\columnwidth]{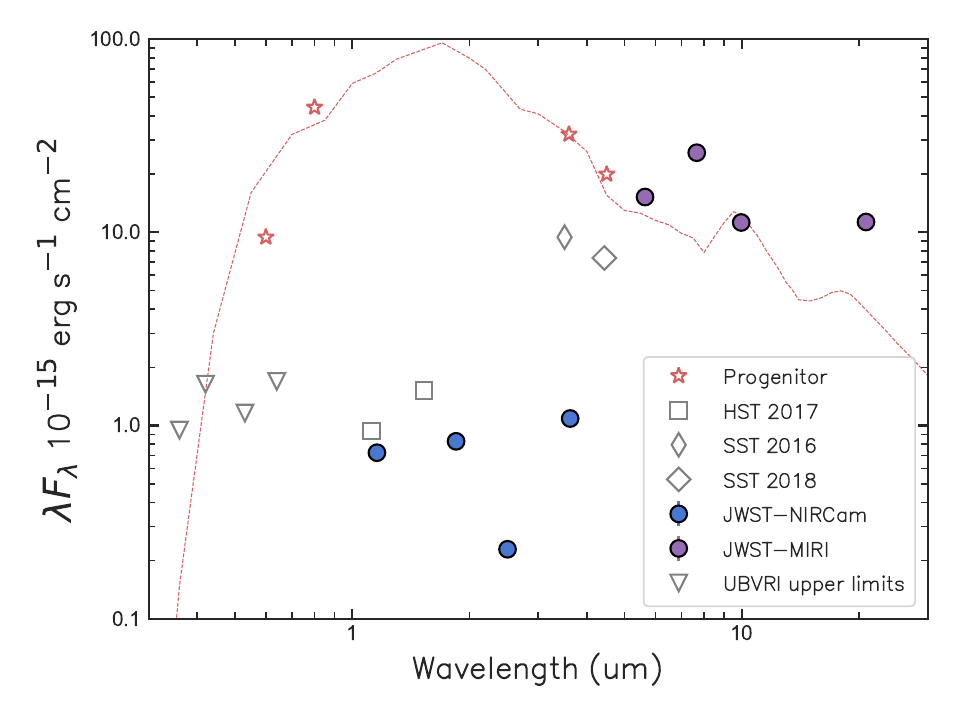}
    \caption{SED of BH1. The JWST data are shown by the filled circles, while archival data is shown by open gray symbols. SST+HST taken from \cite{basinger2021bh1}, upper limits and progenitor photometry taken from \citet{adams2017search}. We also plot our best fitting {\tt DUSTY} model for the progenitor. See Section \ref{sec:lbol} for more detail on the model assumptions. We note the F250M flux is likely inaccurate, see Appendix \ref{sec:appendix}.}
    \label{fig:SED}
\end{figure}

The newly observed SED is significantly fainter and redder than the progenitor star, peaking at 7.7$\mu$m. The peak in emission at infrared wavelengths is indicative of a cooler component to the system, such as a dust shell or previously ejected envelope. We note that the emission profile is narrower than a blackbody and does not appear consistent with silicate dust emission typically observed around RSGs \citep{ohnaka2008spatially}. In this case, the strong F770M emission causes the SED to more closely resemble Case C Polycyclic Aromatic Hydrocarbon (PAH) molecule emission. PAH emission is observed in many astrophysical settings, including around post-asymptotic giant branch (AGB) stars \citep{peeters2002pah} and binary systems (e.g. TU Tau) where a hot companion provides a UV radiation field that illuminates the carbon dust \citep{smolders2010when,boersma2006tutau}. This particular emission profile is discussed further in Section \ref{sec:PAH}.  

We also note that the flux between the NIRCam F118M and F250M filters appears to fall off at a gradient steeper than a blackbody. After comparing the SED to a number of other stars in the field we see this dip in flux in the F250M filter is ubiquitous, implying a flux calibration error in this filter. We therefore advise the flux here is likely an underestimate by a factor of $\sim$ 3, see Appendix \ref{sec:appendix}.

Given their similar wavelengths, it is possible to compare the NIRCam fluxes to the HST and Spitzer Space Telescope (SST) data taken between 2017 - 2018. The fluxes at $\sim$ 1$\mu$m have remained almost constant. The flux at 3.5$\mu$m has dropped significantly, by a factor of $\sim$ 6 compared to the SST data taken in 2016. 

\subsection{Luminosity}\label{sec:lbol}
We now compare the luminosity of the progenitor system to the luminosity we currently observe. Since there are no pre-disappearance constraints on the progenitor flux long ward of 4.5$\mu$m the bolometric luminosity can only be inferred by model fitting, rather than the model-independent method of integrating directly under the observed SED \citep[as in][]{davies2018humphreys, mcdonald2022red}. To estimate the progenitor luminosity we fit {\tt DUSTY} models to the pre-disappearance HST and SST data points. The {\tt DUSTY} models used follow the same setup as described in \citet{beasor2016evolution,beasor2018evolution,beasoe2020new}. We opt for silicate-rich dust as specified by \citet{draine1984optical} and a constant grain size of 0.3$\mu$m \citep[e.g.][]{smith2001asymmetric,scicluna2015large}. We assume a steady state wind such that the density distribution of the dust shell falls off as $r^{-2}$. For consistency, we use the same distances and extinction corrections as done by \cite[][$D$=7.7 Mpc, $E(B-V)$ = 0.303]{basinger2021bh1}, finding a luminosity of \logl\ = 5.3 - 5.5\footnote{Given the pre-explosion photometry at $\lambda$ $<$ 2$\mu$m was likely contaminated by the two other sources identified at the location, we also esimate the pre-disappearance luminosity using only the Spitzer IRAC1 flux and the $BC_{\rm IRAC1}$ from \cite{mcdonald2022red}, also finding a luminosity of \logl=5.5.}. This luminosity range is consistent with the results of \citet{adams2017search} and \cite{basinger2021bh1}.  

To determine a luminosity from the JWST SED we simply integrate under the observed data points, since the SED is more complete, yielding a luminosity of \logl\ = 4.7 $\sim$ 0.2, where the error on luminosity is predominantly driven by the uncertainty on the distance. To ensure no flux is lost at shorter wavelengths, we also estimate the luminosity by bootstrapping a 3000K blackbody to the shortest NIRCam flux, finding this only affects the luminosity by 0.01 dex. We repeat this exercise at longer wavelengths using a cold blackbody at 400K and the longest wavelength MIRI data point, finding the potential missing flux accounts for 0.05 dex. As such, we find the luminosity of BH1 is now between 15\% - 25\% of its progenitor. We note that the near-IR portion of the SED contributes very little to the total bolometric luminosity of the system and so the luminosity reported is insensitive to any issues with the F250M filter (see Appendix \ref{sec:appendix}). 

\subsection{Exploring SED fits}
We now attempt to fit the SED of BH1 with various {\tt DUSTY} models for different outburst scenarios\footnote{Note that the F250M filter is excluded from all fits, see Appendix \ref{sec:appendix}}. 

\subsubsection{A fast outburst}
It is clear from the redness of the SED alone that BH1 is not a normal RSG undergoing quiescent mass-loss \citep[see e.g.][]{beasor2020mass, beasor2022dersgs}. \citet{basinger2021bh1} demonstrated that to hide a star, only a moderate ejecta mass (0.1 - 1\msun, see Fig. 6 within) would be required. If we follow the basic assumptions laid out by \citet{adams2017search} and \citet{basinger2021bh1} and assume an outburst is caused by a mass-loss episode with a velocity of $\sim$200 km/s, over the last 14 years the dust could have traveled up to 600 AU.  Given the location of the inner dust boundary, $R_{\rm in}$, for an RSG is at approximately 1 AU, we created a grid of {\tt DUSTY} models where the outer edge of the wind, $R_{\rm out}$, is truncated at 600$R_{\rm in}$. 

Aside from changing the outer edge of the dust shell, $R_{\rm out}$, these models follow the same setup as those presented in \citet{beasor2016evolution}. We compute grids at 3 different effective temperatures ($T_{\rm eff}$ = 3400, 3600, 3800K) using MARCS stellar atmosphere models \citep{gustafsson2008grid} with inner dust temperatures ($T_{\rm in}$) from 100 - 1200K in steps of 100K, and optical depths ($\tau_{\rm V}$) from 0 to 50 in steps of 0.2. We found that whichever grid was in use, it was difficult to fit both the NIR emission and the mid-IR emission simultaneously. We also tried a number of fits with only the mid-IR data, and again it was not possible to find a good fit using our oxygen-rich models to the unusual shape of the mid-IR emission. In Fig. \ref{fig:dusty600} we show the best fitting model from the \teff\ = 3400K grid. Even at a higher optical depth than is generally seen for most RSGs \citep{beasor2022dersgs} it is not possible to fit the observed SED. For this model we find the total dust mass is approximately 10$^{-3}$ \msun. 

\begin{figure}
    \centering
    \includegraphics[width=\columnwidth]{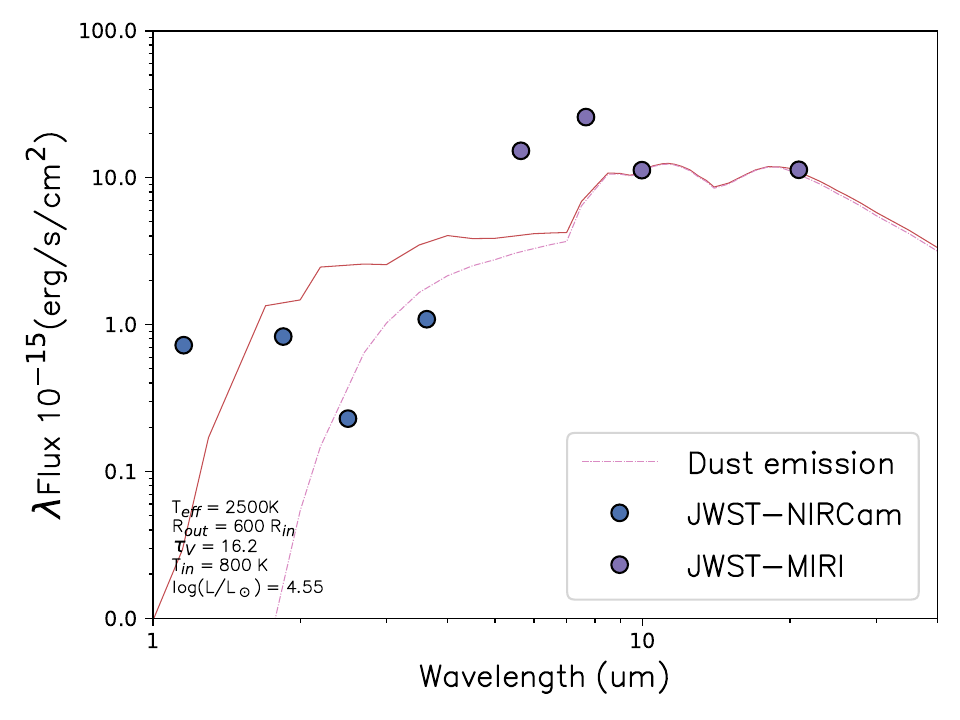}
    \includegraphics[width=\columnwidth]{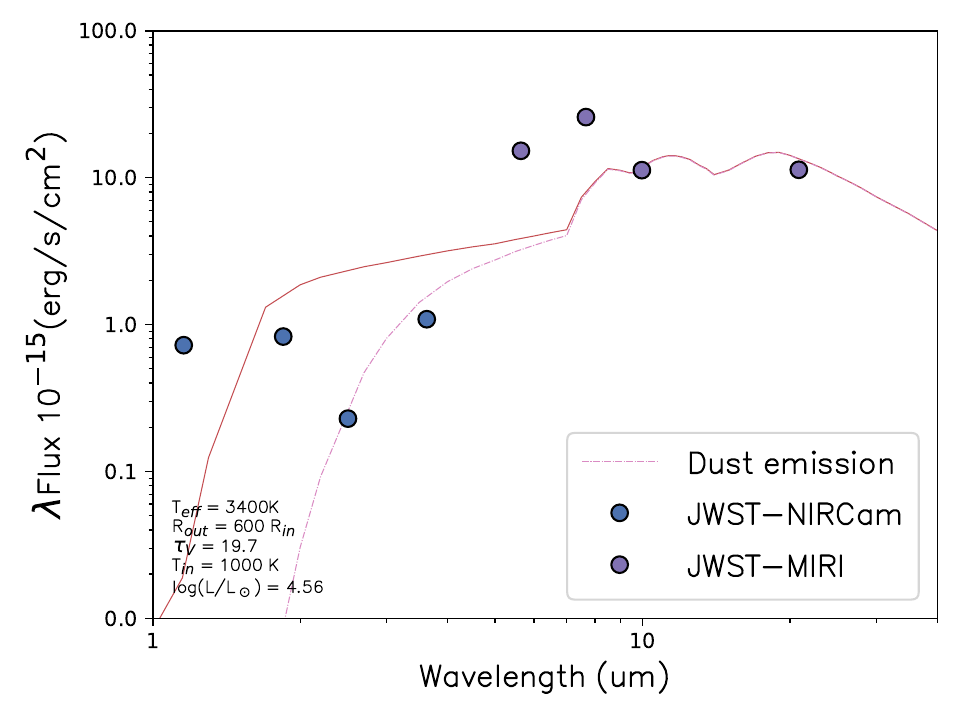}
    \caption{Best fitting {\tt DUSTY} models where the outer radius of the wind is truncated at 600$R_{\rm in}$. {\it Top panel:} Best fit using a 2500K input SED. {\it Bottom panel:} Best fit using a 3400K SED. }
    \label{fig:dusty600}
\end{figure}

\subsubsection{Superwind}
Dust formation is not necessarily the only way to reduce optical flux for RSGs. A 'superwind' is a phase that has been suggested in order to explain early time SN lightcurves \citep{forster2018delay} and has been suggested as an explanation for the extremely red progenitor of SN 2023ixf \citep[see Fig. 1 within][]{jencson2023ixf}. In a superwind, the mass-loss rates are presumed to increase to 10$^{-3}$\msunyr\, --- an order of magnitude lower than the outburst scenario but still a factor of 10-100 times higher than quiescent RSG winds \citep{beasor2020mass}. 

\citet{davies2022superwind} investigated the effect a superwind phase would have on the observed spectra of an RSG. Once the outer superwind reaches $3R_*$, the appearance of the RSG changes dramatically. At this point, the photosphere is inside the wind rather than at the surface of the star, and as such, the photosphere expands as the wind expands. Assuming the effective temperature of the star scales with $1/\sqrt{R}$, the photosphere of the star must now cool (and hence expand) to remain in radiative equilibrium. This cooling causes a shift in the location of the SED peak as well as extra molecular absorption from the CO and CN bands in the J and H bands, and the saturation of the TiO bands at optical wavelengths. In combination, these effects would cause the star to appear fainter and redder, but may not be accompanied by large amounts of dust production. Instead, the tell-tale sign would be molecular emission at $\lambda > 2\mu$m such as from the CO bandhead at $4\mu$m and SiO emission at $8 - 10\mu$m (see Fig. 3).

At some point, it is likely that dust would begin to form in a superwind. In Fig. \ref{fig:superwind_1} we show the SED of a superwind accompanied by a dusty component (here the dusty component is taken from the best-fitting progenitor model). By-eye, this seems to fit the data points reasonably well. To investigate this further we input the superwind SED model into {\tt DUSTY}. If, following \citet{forster2018delay} and \citet{davies2022superwind}, we assume the superwind has a velocity of 10km/s in 14 years it likely will only have travelled a short distance from the surface of the star. As such, we clip $R_{\rm out}$ at 2$R_{\rm in}$. Figure \ref{fig:superwind_2} shows the best fitting superwind model after being input into {\tt DUSTY}. While this model does do a better job of fitting the mid-IR flux, though the mid-IR emission feature does appear to be broader than the observed emission, there is a clear discrepancy at shorter wavelengths.

\begin{figure}
    \centering
    \includegraphics[width=\columnwidth]{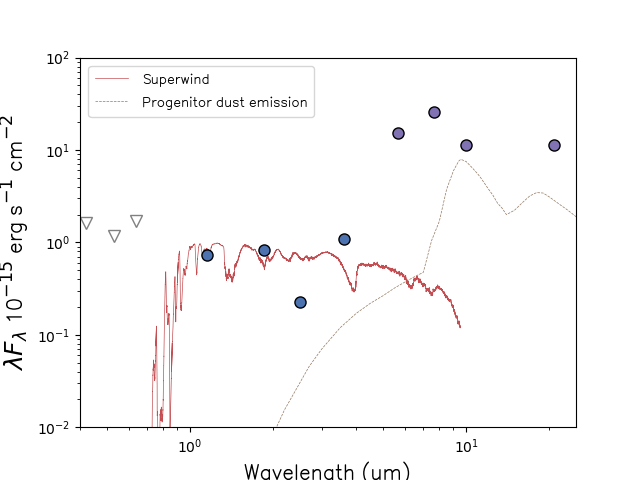}
    \caption{A superwind model for an outer radius of 20$R_{*}$. We show the JWST data in with filled circles, as well as the dust emission from the best-fitting progenitor {\tt DUSTY} model (see Section \ref{sec:lbol}).  }
    \label{fig:superwind_1}
\end{figure}

\begin{figure}
    \centering
    \includegraphics[width=\columnwidth]{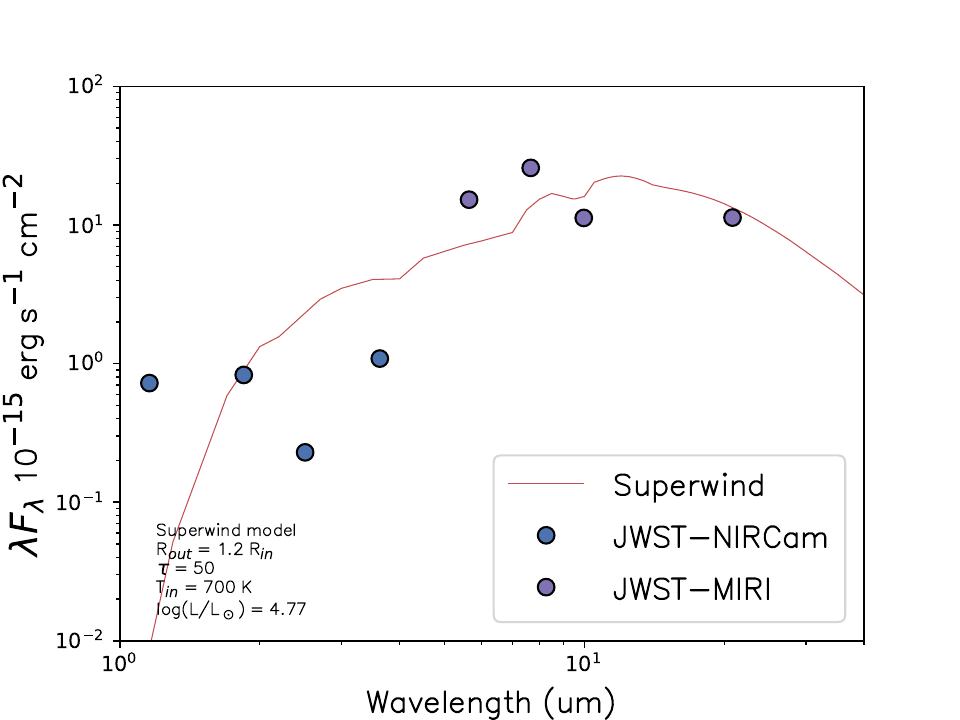}
    \caption{Same as Fig. \ref{fig:superwind_1}, but after the superwind model has been processed through {\tt DUSTY}. }
    \label{fig:superwind_2}
\end{figure}

\subsection{PAH emission}\label{sec:PAH}
We now look at how closely the mid-IR SED resembles PAH emission. As described by \cite{peeters2002pah}, there are a number of distinct varieties of PAH emission. Case C PAH features, characterised by strong emission in the 7-8$\mu$m region are most commonly observed when the dust is ionized \cite[][]{peeters2002pah,draine2021excitation}. \cite{smolders2010when} show that the centroid of the 7.7$\mu$m emission shifts to bluer wavelengths for hotter central stars.  

In Figure \ref{fig:PAH} we show a model Case C PAH spectrum added to a 500K blackbody. We also plot the location of the JWST fluxes in this region. Clearly, this model does a better job of fitting the mid-IR emission than the silicate dust models. We discuss the presence of PAH emission and the potential implications in Section \ref{sec:discussion}. 

\begin{figure}
    \centering
    \includegraphics[width=\columnwidth]{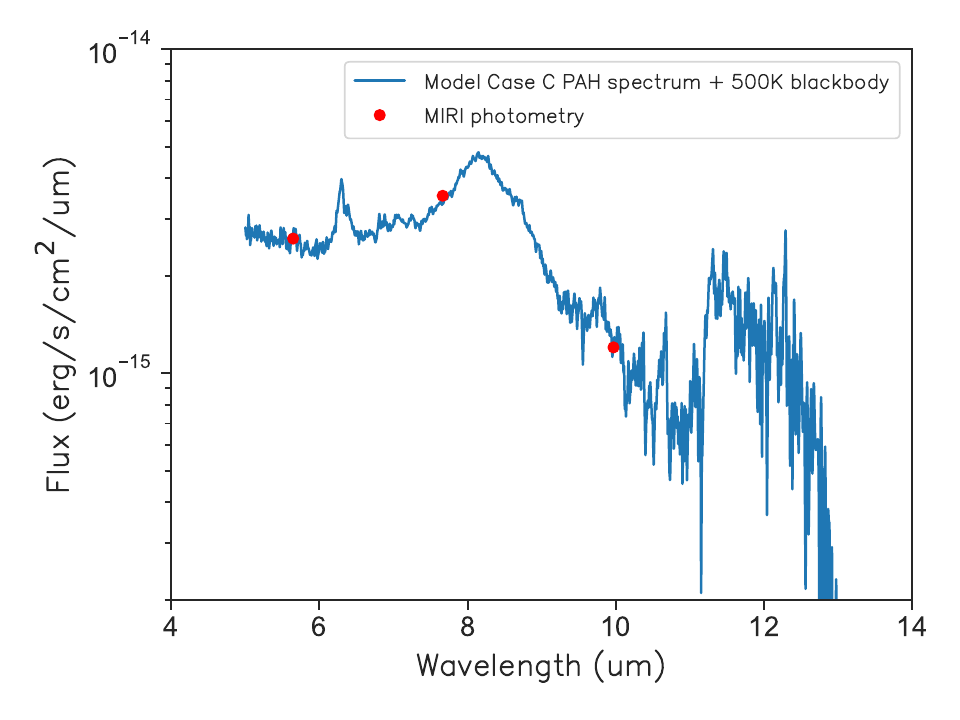}
    \caption{Case C PAH emission profile added to a 500K blackbody. We match the spectrum to the fluxes of the JWST data in this wavelength range. }
    \label{fig:PAH}
\end{figure}

\section{Discussion}\label{sec:discussion}

The new JWST observations presented above do not match either straightforward prediction that would have decisively determined the nature of N6946-BH1. Namely, the source did not disappear as expected for a massive star that has collapsed directly to a BH, and it did not remain at the same bolometric luminosity as expected for a star that survives unchanged except for being surrounded by dust. Instead, the observed result is in between these extremes, where a very luminous IR source persists, although not quite as luminous as the source identified by \citet{gerke2015search} in the pre-disappearance data. 
This ambiguous outcome, plus the multiple sources coincident with the progenitor and the unexpected PAH emission, mean that the situation must be more complicated than previously envisioned.  We now discuss modified scenarios that might explain the fading of BH1 at optical wavelengths, and the remaining mid-IR flux. We also discuss some potential future observations, other than simply waiting for decades, that might help choose between these two.

\subsection{A failed supernova with persistent accretion}\label{sec:disccusion_fsn}

One possibility for a massive star that fades suddenly is that it has imploded; i.e. it has collapsed to a BH without a bright SN display. \citet{kochanek2008survey} designed an observing experiment to monitor nearby galaxies to search for such events, and NGC6946-BH is the first compelling candidate \citep{gerke2015search,adams2017search}.  It has been 14 years since the source faded, and it has not yet re-brightened at visual wavelengths.  On the other hand, a source at the same position has persisted in the near-IR, and has in fact stayed at about the same mid-IR flux level up until the present epoch, as the new JWST images show.

In order for a failed SN to plausibly account for N6946-BH1, there needs to be an explanation for why the source has remained so luminous for so long after collapse to a BH. Unfortunately, theoretical predictions for what a failed SN event may look like in the years after collapse are not yet clear enough to provide testable predictions. Models do seem to predict some level of lingering post-implosion IR luminosity caused by the BH fallback accretion luminosity being absorbed and re-emitted by the dust shell/ejected envelope, but the luminosity can vary greatly.
\citet{perna2014disk} use {\tt MESA} to investigate the fate of fallback matter around compact objects, predicting BH accretion luminosity that ranges from zero up to the same luminosity as the progenitor. This accretion could last anywhere from a few minutes to thousands of years and is strongly dependent on how much of the envelope is ejected and what the remaining mass and angular momentum distribution is for the not-ejected envelope. Armed with only optical-to-IR photometry, this hypothesis is therefore difficult to disprove or verify.  If the mass and angular momentum distribution are tuned the right way, long-lasting accretion to a BH could potentially supply the observed luminosity of N6946-BH1.

There are, however, a few difficulties with the failed SN scenario. While fallback accretion can provide a wide range of luminosity, as noted above, there are limits on how this behaves with time.  For example, most models have a high fallback accretion luminosity on short free-fall timescales, and then a lower luminosity that can persist as the material in the loosely bound envelope with sufficient angular momentum forms a long-lived accretion disk. In this phase, however, the accretion luminosity is generally expected to drop with time as $L \propto t^{-4/3}$ \citep{perna2014disk}, which seems to be contradicted by the relatively constant IR luminosity of N6946-BH1.  In order for fallback accretion to match observations, and to simultaneously eject the right amount of mass to obscure the star and provide the observed transient event, the portion of the stellar envelope that is not successfully ejected may need a fine-tuned density and angular momentum profile, so further work is needed.

Additionally, it is unclear if the observed outburst that preceded the disappearance is consistent with expectations for a failed SN.  \citet{lovegrove2013very} discussed outbursts that may arise in a failed SN because of the hydrodynamic response of the envelope to the loss of rest mass as neutrinos escape the core \citep{nad80}.  In general, the expected transients have a peak luminosity in the right range for the outburst of N6946-BH1, but tend to last longer.  \citet{lovegrove2013very} showed a model light curve for a 15 $M_{\odot}$ failed SN transient that remained bright for over 500 days, whereas the observed BH1 transient lasted only about 100-200 days \citep{adams2017search}.  Moreover, \citet{lovegrove2013very} noted that it tends to be harder for the shock to reach the surface in a more massive 25 $M_{\odot}$ RSG model, since the shock must propagate out through a more massive core and the mass lost via neutrinos tends to be less. 

Recent work presented by \citet{antoni2023failed} suggest the observational signature from a failed SN would differ from the \citet{lovegrove2013very} models. Most notably, the authors suggest a failed SN from a collapsing 25$M_{\odot}$ would be significantly brighter. In the new 3D models, the convective envelope cannot collapse spherically onto the core of the star, creating a disk-like structure that generates outflows and ultimately powers a more luminous transient. \citet{antoni2023failed} demonstrate that the transient created by a collapsing 25\msun\ star would be significantly brighter than the observed outburst seen for N6946-BH1 (see Fig. 13 within). 

Finally, it is unclear if PAH emission is consistent with the failed SN case.  As we noted above, the mid-IR SED from new JWST photometry seems most consistent with emission from a dust shell that also has a strong 7.7 $\mu$m PAH emission band. This emission generally requires dense molecular gas and dust that is exposed to strong near-UV radiation.  A strongly accreting BH would produce most of its luminosity in X-rays and far-UV.  No X-rays were detected from BH1 \citep{basinger2021bh1}, but perhaps these were absorbed by the ejected envelope.  However, one might expect such high-energy photons to mostly dissociate the molecular gas and small grains, leaving mostly the more resilient large dust grains. Indeed, the 7.7$\mu$m PAH emission around active galactic nuclei (AGN) appears weaker than the emission seen in AGB stars since the harsh radiation destroys smaller PAHs, while the 11.3$\mu$m feature is able to survive \citep{xie2022ionization}. Again, additional work on the radiative transfer in this scenario is needed to determine if the likely PAH emission is a problem for the failed SN model.

\subsection{A non-terminal outburst and dust shell}
As noted above, the strong IR source that persists at the position of NGC~6946-BH1 motivates other potential explanations, besides a failed SN.  Since the fading of this source was preceded by a luminous outburst, we consider non-terminal stellar outbursts that may eject a dusty obscuring envelope as a possible alternative.  

Observations have shown that non-terminal stellar outbursts are actually quite common and do eject substantial shells that can obscure the star. Various types of non-terminal or low-energy transients, including luminous red novae \cite[LRNe;][]{howitt2020luminous, pastorello2019LRN,karambelkar2023lrn}, intermediate-luminosity optical transients (ILOTs), luminous blue variable (LBV) eruptions and SN impostors \citep{smith11lbv}, etc., have been studied and have often been associated with stellar mergers or sudden common envelope events in binary systems \citep{blagorodnova2017common, macleod2017common, metzger2017shock,pejcha2017spiral,kashisoker17,smith2016massive,smith18}.   Historically, LBVs like the 1600 AD eruption of P Cygni \citep{smith2006infra} and the 19th century eruption of $\eta$ Carinae \citep{smith2011etacar} were followed by many decades when the stars faded at visual wavelengths.  More recently, stellar merger events like V838~Mon \citep{munari2002v838, bond2003v838} and V1309 Sco \citep{shara2010v1309sco,mason2010v1309sco,ste2011v1309sco} are dramatic, well-studied examples that left dust-obscured sources after their outbursts. A merger or common envelope event such as this in a moderately massive star may provide a suitable explanation for N6946-BH1.  

In fact, \citeauthor{kashisoker17} proposed a specific model wherein N6946-BH1 was the result of an ILOT, where the merger product was obscured by a dusty torus ejected in the merger, and where we are viewing it from a low latitude near the equatorial plane of that system.  Here we discuss this idea in the context of new questions raised by our JWST observations.

In 2009, N6946-BH1 had an outburst when it briefly increased its luminosity by about a factor of 10 for around 100-200 days, before fading by many magnitudes \citep{adams2017search}.  While we noted above that this behavior is not too different from some expectations for models of failed SNe, it is also important to realize that this matches quite well the observed behavior of a number of types of non-terminal transients mentioned above.  In fact, this ambiguity is unavoidable; as noted by \citet{lovegrove2013very}, the observable transients that they predict are difficult to distinguish from LRNe and other stellar outbursts.  The physical reason for the similarity is that both outbursts from failed SNe and stellar mergers result from the incomplete ejection of the H envelope in a weak shock, and the resulting light arises from recombination emission as the ejecta expand and cool, forming dust and molecules.  Thus, the outburst itself is consistent with either model.  The prelude and aftermath can be different in the two cases, of course.

In terms of the aftermath, the main question posed by the JWST observations is why the lingering mid-IR source would be less luminous than the progenitor star. We noted that the integrated bolometric luminosity of the source currently at the position of N6946-BH1 is about \logl=4.7, while still a luminous source, this is only 13 -- 25\% of the total luminosity estimated for the pre-outburst source \citep[See Section \ref{sec:lbol} and ][]{gerke2015search}. If a star survives, \citet{kochanek2012absorption} suggests that any flux lost at shorter wavelengths is re-emitted at longer wavelengths by the obscuring dust shell, and the total luminosity of the system should be conserved.  A scenario where the 2009 outburst was a stellar merger provides two natural answers to this; either may work, but both are expected.  

First, the IR luminosity we detect for the obscured source would only match the luminosity of the embedded surviving star if the dust shell is spherical. In a stellar merger event, the assumption of spherical symmetry for the ejected shell is not only an oversimplification, but is almost certainly an incorrect assumption.  In a stellar merger event, much of the ejected mass is necessarily in a flatted equatorial distribution, since angular momentum must be shed in order to allow the merger to occur.  If the dust shell is not spherical, then the dusty torus only re-radiates a portion of the source's luminosity.   This fraction of the total stellar luminosity that is reradiated in the mid-IR is essentially the fraction of the solid angle intercepted by the torus as seen by the central star, which may plausibly be 10-20\%.   This idea was discussed in detail by \citeauthor{kashisoker17}, who adopted a covering fraction for the torus of about 20\% and argued that this can explain N6946-BH1 very well.  \citeauthor{kashisoker17} also predicted that N6946-BH1 would be a luminous source at longer wavelengths than were available at that time, which is in good agreement with the observations we present here. In order for this scenario to work, we must be observing the system from a vantage point close to the equatorial plane, looking through the torus. Given a population of such events, the fraction that will be observed favorably through the torus corresponds to the same fraction as the fraction of absorbed and re-rated IR luminosity (i.e. the solid angle covering fraction of the torus).  Thus, a small but non-zero fraction of non-terminal transients should look like N6946-BH1.  This is in reasonable agreement with the observed diversity of such events.

A second reason why the late-time luminosity we see now might be less than before disappearance, even if the star survived, is that the pre-outburst luminosity was probably elevated compared to the star's normal luminosity.  This is expected in the scenario where the 2009 outburst was caused by a merger or common envelope event.  The source detected in 2007, a year and a half before its outburst, was shown to be consistent with a red star that had $\log(L/L_{\odot}) = 5.3-5.5$ \citep{gerke2015search}.  This was interpreted as an RSG progenitor with a high initial mass of  $20-25\,M_{\odot}$.  However, if the pre-outburst star was actually in an elevated state, then this translation to initial mass would be erroneous. 

In order for a stellar merger to occur, there needs to be a relatively long-lasting (i.e. years to decades) inspiral phase when the orbital separation shrinks as mass and angular momentum are shed by the system, and this can produce a substantially higher luminosity than the stars make on their own. For example, the LRN eruption of V1309 Sco appeared to brighten by almost an order of magnitude prior to merging in 2009 \citep{tylenda2011v1309sco}. \citet{pejcha2017spiral} propose a model to explain the pre-merger lightcurve of V1309 Sco. They suggest the star underwent a `death-spiral' resulting from a runaway loss of mass and angular momentum from the outer Lagrange point, and the shock heating of the disk strongly increased the emergent luminosity. Similarly, the much more massive system $\eta$ Car was increasing in brightness for decades preceding its merger event in the 1840s \citep{sf11,smith18}.  The pre-outburst behavior of N6946-BH1 shows some features consistent with such an inspiral phase.  For example, the pre-outburst lightcurve presented in \cite{adams2017search} shows that between 2005 - 2009 the progenitor brightened by almost a factor of 10 in the Spitzer 3.5 $\mu$m and 4.5$\mu$m filters. Similar behaviour was seen for V1309 Sco. At visual wavelengths, the star brightened slowly by almost a factor of 2 from 1999 to 2005 \citep{adams2017search}, and during this time the color was significantly bluer than the red color observed in 2007, from which the 20-25 $M_{\odot}$ RSG progenitor was inferred \citep{gerke2015search}.   Then the source's blue photometry actually showed a strong drop in flux from 2005 until its outburst in 2009, coinciding with the time when the IR flux dramatically increased. In fact, \cite{adams2017search} found the best fitting models progenitor models for the 2005 data were hotter stars, with their model P1 having a best fit $T_{\rm eff}$ = 6700$^{+1200}_{-990}$ K. This may suggest that the source appeared red immediately before outburst because it was reddened by dust, not necessarily because it was a very cool star \citep[see e.g.][]{humphreys2019bh1}. If the outburst and subsequent fading of BH1 was due to a merger, it is consistent with a merger involving a YSG, rather than an RSG progenitor (for which no merger event has been observed). A similar dip was seen in both observations and model light curves of the merger event V1309 Sco, where the pre-outburst dip is caused by a rapid increase in extinction from dust in the outflowing death spiral right before a merger. 

Finally, a common envelope or merger event could also help account for the potential presence of PAH emission in the mid-IR, which we inferred above from the shape of the SED using JWST photometry. Often, PAH emission is seen for AGB stars in binary systems since the dust is illuminated by a UV field from a hot companion \citep[e.g. TU Tau, ][]{boersma2006tutau}. It has also been observed around LBVs such as HD168625 \citep{umana2010spitz} and R71 \citep{niyogi2014r71}, as well as yellow hypergiant HR5171A \citep{ches2014hr}. Interestingly, PAH features have also been observed around SN imposter NGC 300 OT \citep{prieto2009300} and intermediate luminous red transient (ILRT) AT 2019abn (Rose et al. in prep). The PAH emission requires a dense, cool, dusty molecular envelope that is irradiated by near-UV flux (and as such, may not require the surrounding dust to be carbon-rich rather than silicate-rich).\footnote{The amount of UV flux may vary with the initial mass of the system, and so for lower mass mergers such as V1309 Sco and V4332 Sgr there may not be enough UV flux present to allow PAH features to form. Other mergers for which no PAH features have been confirmed, the observations were often taken a long time after the event and so any PAH features may have already faded.} In the case of a binary merger or common envelope event, that internal UV source may be a hotter companion star or the stripped He core of the star from which the envelope was ejected.

\subsection{Implications and future tests}
The currently available data means it is difficult to decipher between the outburst and accreting BH scenarios. We may be able to break this degeneracy with JWST MIRI spectra which would show whether or not PAH emission is present. As discussed in \ref{sec:disccusion_fsn} it seems unlikely molecular gas and dust grains would survive being irradiated the high energy photons emitted by an accreting BH, though this likely needs to be further explored in theoretical works.

\section{Summary}
We report new JWST near- and mid-IR detections of the source that remains at the location of BH1. Below we summarise the main results of this study. 
\begin{itemize}

\item NIRCam reveals BH1 was likely a blend of at least 3 stars within a triangular asterism that were not resolved in previous data sets from e.g. HST and SST.

\item A luminous source remains in the IR, with a mid-IR flux that is comparable to the presumed mid-IR flux of the progenitor. The bolometric luminosity of this persistent source is only 13\% - 25\% of the pre-disappearance source. This reduction means that we cannot attribute the dip in optical flux of BH1 to a simple, spherical outburst of dust. It is however consistent with a merger scenario resulting in a highly aspherical dusty torus in the equatorial plane as suggested by \citet{kashisoker17}. In the merger scenario the pre-disappearance luminosity inferred from HST and SST may be elevated far beyond the quiescent \lbol\ of the system due to the star shedding angular momentum and mass during the death-spiral \citep[as suggested for V1309 Sco, ][]{pejcha2017spiral}, implying a lower luminosity (and hence initial mass) progenitor. If BH1 has undergone a non-terminal merger event, it could take decades for the dust to clear and for the luminosity of the system to return to its pre-elevation state.

\item The shape of the mid-IR SED does not appear to be consistent with normal, RSG mass loss. Instead, the narrow emission profile appears consistent with PAH features commonly seen around AGB stars, as well as LBVs, ILRTs and a SN imposter. PAH emission implies dust is being irradiated by a hot, UV-bright source. We suggest that this is best explained by a common envelope merger event, which would also explain the 2009 outburst of BH1. 

\end{itemize}

\begin{acknowledgements}
    The authors would like to thank Ryan Lau for useful discussions and for providing the model PAH emission spectrum. We would also like to thank the anonymous referee who provided comments which improved the paper. 

    Time-domain research by D.J.S.\, G.H. and J.P. is supported by NSF grants AST-1821987, 1813466, 1908972, 2108032, and 2308181, and by the Heising-Simons Foundation under grant \#2020-1864.
\end{acknowledgements}

\vspace{5mm}
\facilities{JWST(MIRI and NIRCam)}




\appendix
\section{NIRCam flux discrepancy}\label{sec:appendix}
We find that the flux from the NIRCAM-F250M filter appears to drop off at a rate steeper than a blackbody for BH1. While inconsequential to the conclusions of this work (because the source is so faint at these wavelengths anyway), it is necessary to assess whether this feature is real (e.g. caused by either an absorption line at 2.5$\mu$m or an emission line at 1.8$\mu$m) or whether it is an error in flux calibration. 

To check this, we retrieved photometry for 6 random field stars that should all have roughly blackbody-shaped SEDs, see Fig. \ref{fig:f250}.  We find that in each case the F250M flux appears too low as compared to the typical SED shape of a stellar photosphere that passes through the other filters. Since the feature is ubiquitous, we can assume the F250M photometry likely suffers from a calibration error, and that the flux of BH1 is likely higher than we have derived here. We are not currently able to determine the cause of the dip, so instead we tentatively suggest that the F250M flux likely requires a correction factor of $\approx$2.6.

\begin{figure*}
    \centering
    \includegraphics[width=10cm]{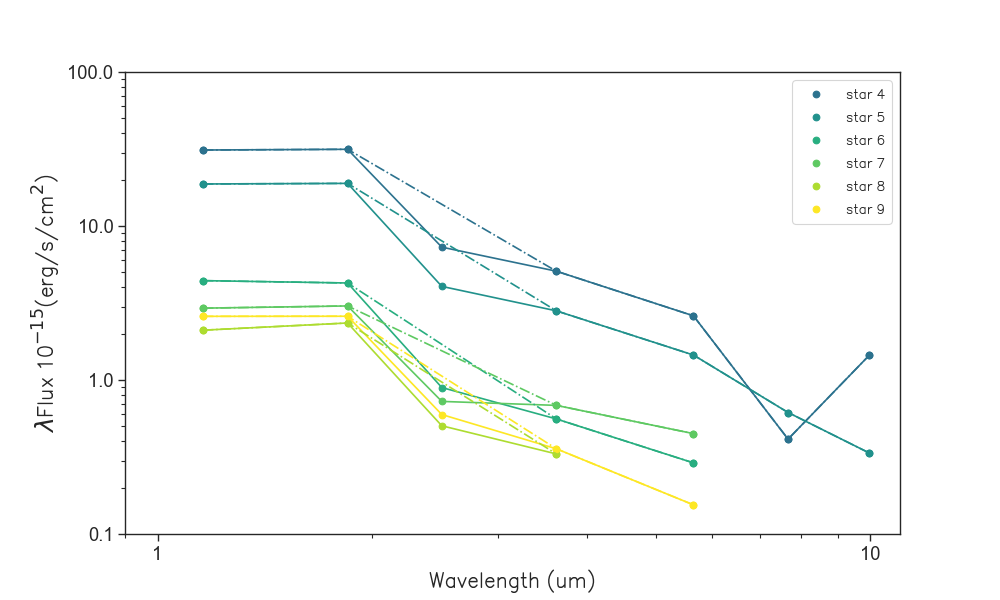}
    \caption{SEDs of the 6 field stars used to asses the F250M filter flux. The solid lines show the observed SED, while the dashed lines show the approximate expected flux were the SEDs to follow a blackbody curve.}
    \label{fig:f250}
\end{figure*}



\bibliography{references}{}

\begin{thebibliography}{}
\expandafter\ifx\csname natexlab\endcsname\relax\def\natexlab#1{#1}\fi
\providecommand{\url}[1]{\href{#1}{#1}}
\providecommand{\dodoi}[1]{doi:~\href{http://doi.org/#1}{\nolinkurl{#1}}}
\providecommand{\doeprint}[1]{\href{http://ascl.net/#1}{\nolinkurl{http://ascl.net/#1}}}
\providecommand{\doarXiv}[1]{\href{https://arxiv.org/abs/#1}{\nolinkurl{https://arxiv.org/abs/#1}}}

\bibitem[{{Abbott} {et~al.}(2017){Abbott}, {Abbott}, {Abbott}, {Acernese},
  {Ackley}, {Adams}, {Adams}, {Addesso}, {Adhikari}, {Adya}, {Affeldt},
  {Afrough}, {Agarwal}, {Agathos}, {Agatsuma}, {Aggarwal}, {Aguiar}, {Aiello},
  {Ain}, {Ajith}, {Allen}, {Allen}, {Allocca}, {Altin}, {Amato}, {Ananyeva},
  {Anderson}, {Anderson}, {Antier}, {Appert}, {Arai}, {Araya}, {Areeda},
  {Arnaud}, {Arun}, {Ascenzi}, {Ashton}, {Ast}, {Aston}, {Astone}, {Aufmuth},
  {Aulbert}, {AultONeal}, {Avila-Alvarez}, {Babak}, {Bacon}, {Bader}, {Bae},
  {Baker}, {Baldaccini}, {Ballardin}, {Ballmer}, {Banagiri}, {Barayoga},
  {Barclay}, {Barish}, {Barker}, {Barone}, {Barr}, {Barsotti}, {Barsuglia},
  {Barta}, {Bartlett}, {Bartos}, {Bassiri}, {Basti}, {Batch}, {Baune}, {Bawaj},
  {Bazzan}, {B{\'e}csy}, {Beer}, {Bejger}, {Belahcene}, {Bell}, {Berger},
  {Bergmann}, {Berry}, {Bersanetti}, {Bertolini}, {Betzwieser}, {Bhagwat},
  {Bhandare}, {Bilenko}, {Billingsley}, {Billman}, {Birch}, {Birney},
  {Birnholtz}, {Biscans}, {Bisht}, {Bitossi}, {Biwer}, {Bizouard}, {Blackburn},
  {Blackman}, {Blair}, {Blair}, {Blair}, {Bloemen}, {Bock}, {Bode}, {Boer},
  {Bogaert}, {Bohe}, {Bondu}, {Bonnand}, {Boom}, {Bork}, {Boschi}, {Bose},
  {Bouffanais}, {Bozzi}, {Bradaschia}, {Brady}, {Braginsky}, {Branchesi},
  {Brau}, {Briant}, {Brillet}, {Brinkmann}, {Brisson}, {Brockill}, {Broida},
  {Brooks}, {Brown}, {Brown}, {Brown}, {Brunett}, {Buchanan}, {Buikema},
  {Bulik}, {Bulten}, {Buonanno}, {Buskulic}, {Buy}, {Byer}, {Cabero},
  {Cadonati}, {Cagnoli}, {Cahillane}, {Calder{\'o}n Bustillo}, {Callister},
  {Calloni}, {Camp}, {Canepa}, {Canizares}, {Cannon}, {Cao}, {Cao}, {Capano},
  {Capocasa}, {Carbognani}, {Caride}, {Carney}, {Casanueva Diaz}, {Casentini},
  {Caudill}, {Cavagli{\`a}}, {Cavalier}, {Cavalieri}, {Cella}, {Cepeda},
  {Cerboni Baiardi}, {Cerretani}, {Cesarini}, {Chamberlin}, {Chan}, {Chao},
  {Charlton}, {Chassande-Mottin}, {Chatterjee}, {Chatziioannou}, {Cheeseboro},
  {Chen}, {Chen}, {Cheng}, {Chincarini}, {Chiummo}, {Chmiel}, {Cho}, {Cho},
  {Chow}, {Christensen}, {Chu}, {Chua}, {Chua}, {Chung}, {Chung}, {Ciani},
  {Ciolfi}, {Cirelli}, {Cirone}, {Clara}, {Clark}, {Cleva}, {Cocchieri},
  {Coccia}, {Cohadon}, {Colla}, {Collette}, {Cominsky}, {Constancio}, {Conti},
  {Cooper}, {Corban}, {Corbitt}, {Corley}, {Cornish}, {Corsi}, {Cortese},
  {Costa}, {Coughlin}, {Coughlin}, {Coulon}, {Countryman}, {Couvares}, {Covas},
  {Cowan}, {Coward}, {Cowart}, {Coyne}, {Coyne}, {Creighton}, {Creighton},
  {Cripe}, {Crowder}, {Cullen}, {Cumming}, {Cunningham}, {Cuoco}, {Dal Canton},
  {Danilishin}, {D'Antonio}, {Danzmann}, {Dasgupta}, {Da Silva Costa},
  {Dattilo}, {Dave}, {Davier}, {Davis}, {Daw}, {Day}, {De}, {DeBra}, {Deelman},
  {Degallaix}, {De Laurentis}, {Del{\'e}glise}, {Del Pozzo}, {Denker}, {Dent},
  {Dergachev}, {De Rosa}, {DeRosa}, {DeSalvo}, {Devenson}, {Devine},
  {Dhurandhar}, {D{\'\i}az}, {Di Fiore}, {Di Giovanni}, {Di Girolamo}, {Di
  Lieto}, {Di Pace}, {Di Palma}, {Di Renzo}, {Doctor}, {Dolique}, {Donovan},
  {Dooley}, {Doravari}, {Dorrington}, {Douglas}, {Dovale {\'A}lvarez},
  {Downes}, {Drago}, {Drever}, {Driggers}, {Du}, {Ducrot}, {Duncan}, {Dwyer},
  {Edo}, {Edwards}, {Effler}, {Eggenstein}, {Ehrens}, {Eichholz}, {Eikenberry},
  {Eisenstein}, {Essick}, {Etienne}, {Etzel}, {Evans}, {Evans}, {Factourovich},
  {Fafone}, {Fair}, {Fairhurst}, {Fan}, {Farinon}, {Farr}, {Farr},
  {Fauchon-Jones}, {Favata}, {Fays}, {Fehrmann}, {Feicht}, {Fejer},
  {Fernandez-Galiana}, {Ferrante}, {Ferreira}, {Ferrini}, {Fidecaro}, {Fiori},
  {Fiorucci}, {Fisher}, {Flaminio}, {Fletcher}, {Fong}, {Forsyth}, {Forsyth},
  {Fournier}, {Frasca}, {Frasconi}, {Frei}, {Freise}, {Frey}, {Frey}, {Fries},
  {Fritschel}, {Frolov}, {Fulda}, {Fyffe}, {Gabbard}, {Gabel}, {Gadre},
  {Gaebel}, {Gair}, {Gammaitoni}, {Ganija}, {Gaonkar}, {Garufi}, {Gaudio},
  {Gaur}, {Gayathri}, {Gehrels}, {Gemme}, {Genin}, {Gennai}, {George},
  {George}, {Gergely}, {Germain}, {Ghonge}, {Ghosh}, {Ghosh}, {Ghosh},
  {Giaime}, {Giardina}, {Giazotto}, {Gill}, {Glover}, {Goetz}, {Goetz},
  {Gomes}, {Gonz{\'a}lez}, {Gonzalez Castro}, {Gopakumar}, {Gorodetsky},
  {Gossan}, {Gosselin}, {Gouaty}, {Grado}, {Graef}, {Granata}, {Grant}, {Gras},
  {Gray}, {Greco}, {Green}, {Groot}, {Grote}, {Grunewald}, {Gruning}, {Guidi},
  {Guo}, {Gupta}, {Gupta}, {Gushwa}, {Gustafson}, {Gustafson}, {Hall}, {Hall},
  {Hammond}, {Haney}, {Hanke}, {Hanks}, {Hanna}, {Hannam}, {Hannuksela},
  {Hanson}, {Hardwick}, {Harms}, {Harry}, {Harry}, {Hart}, {Haster},
  {Haughian}, {Healy}, {Heidmann}, {Heintze}, {Heitmann}, {Hello}, {Hemming},
  {Hendry}, {Heng}, {Hennig}, {Henry}, {Heptonstall}, {Heurs}, {Hild}, {Hoak},
  {Hofman}, {Holt}, {Holz}, {Hopkins}, {Horst}, {Hough}, {Houston}, {Howell},
  {Hu}, {Huerta}, {Huet}, {Hughey}, {Husa}, {Huttner}, {Huynh-Dinh}, {Indik},
  {Ingram}, {Inta}, {Intini}, {Isa}, {Isac}, {Isi}, {Iyer}, {Izumi}, {Jacqmin},
  {Jani}, {Jaranowski}, {Jawahar}, {Jim{\'e}nez-Forteza}, {Johnson},
  {Johnson-McDaniel}, {Jones}, {Jones}, {Jonker}, {Ju}, {Junker}, {Kalaghatgi},
  {Kalogera}, {Kandhasamy}, {Kang}, {Kanner}, {Karki}, {Karvinen}, {Kasprzack},
  {Katolik}, {Katsavounidis}, {Katzman}, {Kaufer}, {Kawabe},
  {K{\'e}f{\'e}lian}, {Keitel}, {Kemball}, {Kennedy}, {Kent}, {Key}, {Khalili},
  {Khan}, {Khan}, {Khan}, {Khazanov}, {Kijbunchoo}, {Kim}, {Kim}, {Kim}, {Kim},
  {Kim}, {Kimbrell}, {King}, {King}, {Kirchhoff}, {Kissel}, {Kleybolte},
  {Klimenko}, {Koch}, {Koehlenbeck}, {Koley}, {Kondrashov}, {Kontos},
  {Korobko}, {Korth}, {Kowalska}, {Kozak}, {Kr{\"a}mer}, {Kringel}, {Krishnan},
  {Kr{\'o}lak}, {Kuehn}, {Kumar}, {Kumar}, {Kumar}, {Kuo}, {Kutynia}, {Kwang},
  {Lackey}, {Lai}, {Landry}, {Lang}, {Lange}, {Lantz}, {Lanza},
  {Lartaux-Vollard}, {Lasky}, {Laxen}, {Lazzarini}, {Lazzaro}, {Leaci},
  {Leavey}, {Lee}, {Lee}, {Lee}, {Lee}, {Lee}, {Lehmann}, {Lenon}, {Leonardi},
  {Leroy}, {Letendre}, {Levin}, {Li}, {Libson}, {Littenberg}, {Liu}, {Lo},
  {Lockerbie}, {London}, {Lord}, {Lorenzini}, {Loriette}, {Lormand}, {Losurdo},
  {Lough}, {Lovelace}, {L{\"u}ck}, {Lumaca}, {Lundgren}, {Lynch}, {Ma},
  {Macfoy}, {Machenschalk}, {MacInnis}, {Macleod}, {Maga{\~n}a Hernandez},
  {Maga{\~n}a-Sandoval}, {Maga{\~n}a Zertuche}, {Magee}, {Majorana},
  {Maksimovic}, {Man}, {Mandic}, {Mangano}, {Mansell}, {Manske}, {Mantovani},
  {Marchesoni}, {Marion}, {M{\'a}rka}, {M{\'a}rka}, {Markakis}, {Markosyan},
  {Maros}, {Martelli}, {Martellini}, {Martin}, {Martynov}, {Marx}, {Mason},
  {Masserot}, {Massinger}, {Masso-Reid}, {Mastrogiovanni}, {Matas},
  {Matichard}, {Matone}, {Mavalvala}, {Mayani}, {Mazumder}, {McCarthy},
  {McClelland}, {McCormick}, {McCuller}, {McGuire}, {McIntyre}, {McIver},
  {McManus}, {McRae}, {McWilliams}, {Meacher}, {Meadors}, {Meidam},
  {Mejuto-Villa}, {Melatos}, {Mendell}, {Mercer}, {Merilh}, {Merzougui},
  {Meshkov}, {Messenger}, {Messick}, {Metzdorff}, {Meyers}, {Mezzani}, {Miao},
  {Michel}, {Middleton}, {Mikhailov}, {Milano}, {Miller}, {Miller}, {Miller},
  {Miller}, {Millhouse}, {Minazzoli}, {Minenkov}, {Ming}, {Mishra}, {Mitra},
  {Mitrofanov}, {Mitselmakher}, {Mittleman}, {Moggi}, {Mohan}, {Mohapatra},
  {Montani}, {Moore}, {Moore}, {Moraru}, {Moreno}, {Morriss}, {Mours},
  {Mow-Lowry}, {Mueller}, {Muir}, {Mukherjee}, {Mukherjee}, {Mukherjee},
  {Mukund}, {Mullavey}, {Munch}, {Muniz}, {Murray}, {Napier}, {Nardecchia},
  {Naticchioni}, {Nayak}, {Nelemans}, {Nelson}, {Neri}, {Nery}, {Neunzert},
  {Newport}, {Newton}, {Ng}, {Nguyen}, {Nichols}, {Nielsen}, {Nissanke},
  {Nitz}, {Noack}, {Nocera}, {Nolting}, {Normandin}, {Nuttall}, {Oberling},
  {Ochsner}, {Oelker}, {Ogin}, {Oh}, {Oh}, {Ohme}, {Oliver}, {Oppermann},
  {Oram}, {O'Reilly}, {Ormiston}, {Ortega}, {O'Shaughnessy}, {Ottaway},
  {Overmier}, {Owen}, {Pace}, {Page}, {Page}, {Pai}, {Pai}, {Palamos},
  {Palashov}, {Palomba}, {Pal-Singh}, {Pan}, {Pang}, {Pang}, {Pankow},
  {Pannarale}, {Pant}, {Paoletti}, {Paoli}, {Papa}, {Paris}, {Parker},
  {Pascucci}, {Pasqualetti}, {Passaquieti}, {Passuello}, {Patricelli},
  {Pearlstone}, {Pedraza}, {Pedurand}, {Pekowsky}, {Pele}, {Penn}, {Perez},
  {Perreca}, {Perri}, {Pfeiffer}, {Phelps}, {Piccinni}, {Pichot},
  {Piergiovanni}, {Pierro}, {Pillant}, {Pinard}, {Pinto}, {Pitkin}, {Poggiani},
  {Popolizio}, {Porter}, {Post}, {Powell}, {Prasad}, {Pratt}, {Predoi},
  {Prestegard}, {Prijatelj}, {Principe}, {Privitera}, {Prodi}, {Prokhorov},
  {Puncken}, {Punturo}, {Puppo}, {P{\"u}rrer}, {Qi}, {Qin}, {Qiu}, {Quetschke},
  {Quintero}, {Quitzow-James}, {Raab}, {Rabeling}, {Radkins}, {Raffai}, {Raja},
  {Rajan}, {Rakhmanov}, {Ramirez}, {Rapagnani}, {Raymond}, {Razzano}, {Read},
  {Regimbau}, {Rei}, {Reid}, {Reitze}, {Rew}, {Reyes}, {Ricci}, {Ricker},
  {Rieger}, {Riles}, {Rizzo}, {Robertson}, {Robie}, {Robinet}, {Rocchi},
  {Rolland}, {Rollins}, {Roma}, {Romano}, {Romano}, {Romel}, {Romie},
  {Rosi{\'n}ska}, {Ross}, {Rowan}, {R{\"u}diger}, {Ruggi}, {Ryan}, {Rynge},
  {Sachdev}, {Sadecki}, {Sadeghian}, {Sakellariadou}, {Salconi}, {Saleem},
  {Salemi}, {Samajdar}, {Sammut}, {Sampson}, {Sanchez}, {Sandberg}, {Sandeen},
  {Sanders}, {Sassolas}, {Sathyaprakash}, {Saulson}, {Sauter}, {Savage},
  {Sawadsky}, {Schale}, {Scheuer}, {Schmidt}, {Schmidt}, {Schmidt}, {Schnabel},
  {Schofield}, {Sch{\"o}nbeck}, {Schreiber}, {Schuette}, {Schulte}, {Schutz},
  {Schwalbe}, {Scott}, {Scott}, {Seidel}, {Sellers}, {Sengupta}, {Sentenac},
  {Sequino}, {Sergeev}, {Shaddock}, {Shaffer}, {Shah}, {Shahriar}, {Shao},
  {Shapiro}, {Shawhan}, {Sheperd}, {Shoemaker}, {Shoemaker}, {Siellez},
  {Siemens}, {Sieniawska}, {Sigg}, {Silva}, {Singer}, {Singer}, {Singh},
  {Singh}, {Singhal}, {Sintes}, {Slagmolen}, {Smith}, {Smith}, {Smith}, {Son},
  {Sonnenberg}, {Sorazu}, {Sorrentino}, {Souradeep}, {Spencer}, {Srivastava},
  {Staley}, {Steinke}, {Steinlechner}, {Steinlechner}, {Steinmeyer},
  {Stephens}, {Stevenson}, {Stone}, {Strain}, {Stratta}, {Strigin}, {Sturani},
  {Stuver}, {Summerscales}, {Sun}, {Sunil}, {Sutton}, {Swinkels},
  {Szczepa{\'n}czyk}, {Tacca}, {Talukder}, {Tanner}, {T{\'a}pai}, {Taracchini},
  {Taylor}, {Taylor}, {Theeg}, {Thomas}, {Thomas}, {Thomas}, {Thorne},
  {Thorne}, {Thrane}, {Tiwari}, {Tiwari}, {Tokmakov}, {Toland}, {Tonelli},
  {Tornasi}, {Torrie}, {T{\"o}yr{\"a}}, {Travasso}, {Traylor}, {Trifir{\`o}},
  {Trinastic}, {Tringali}, {Trozzo}, {Tsang}, {Tse}, {Tso}, {Tuyenbayev},
  {Ueno}, {Ugolini}, {Unnikrishnan}, {Urban}, {Usman}, {Vahi}, {Vahlbruch},
  {Vajente}, {Valdes}, {Vallisneri}, {van Bakel}, {van Beuzekom}, {van den
  Brand}, {Van Den Broeck}, {Vander-Hyde}, {van der Schaaf}, {van Heijningen},
  {van Veggel}, {Vardaro}, {Varma}, {Vass}, {Vas{\'u}th}, {Vecchio},
  {Vedovato}, {Veitch}, {Veitch}, {Venkateswara}, {Venugopalan}, {Verkindt},
  {Vetrano}, {Vicer{\'e}}, {Viets}, {Vinciguerra}, {Vine}, {Vinet}, {Vitale},
  {Vo}, {Vocca}, {Vorvick}, {Voss}, {Vousden}, {Vyatchanin}, {Wade}, {Wade},
  {Wade}, {Wald}, {Walet}, {Walker}, {Wallace}, {Walsh}, {Wang}, {Wang},
  {Wang}, {Wang}, {Wang}, {Wang}, {Ward}, {Warner}, {Was}, {Watchi}, {Weaver},
  {Wei}, {Weinert}, {Weinstein}, {Weiss}, {Wen}, {Wessel}, {We{\ss}els},
  {Westphal}, {Wette}, {Whelan}, {Whiting}, {Whittle}, {Williams}, {Williams},
  {Williamson}, {Willis}, {Willke}, {Wimmer}, {Winkler}, {Wipf}, {Wittel},
  {Woan}, {Woehler}, {Wofford}, {Wong}, {Worden}, {Wright}, {Wu}, {Wu}, {Yam},
  {Yamamoto}, {Yancey}, {Yap}, {Yu}, {Yu}, {Yvert}, {Zadro{\.Z}ny}, {Zanolin},
  {Zelenova}, {Zendri}, {Zevin}, {Zhang}, {Zhang}, {Zhang}, {Zhang}, {Zhao},
  {Zhou}, {Zhou}, {Zhu}, {Zimmerman}, {Zucker}, {Zweizig}, {LIGO Scientific},
  \& {Virgo Collaboration}}]{abbott2017gw1}
{Abbott}, B.~P., {Abbott}, R., {Abbott}, T.~D., {et~al.} 2017, \prl, 118,
  221101, \dodoi{10.1103/PhysRevLett.118.221101}

\bibitem[{{Adams} {et~al.}(2017){Adams}, {Kochanek}, {Gerke}, {Stanek}, \&
  {Dai}}]{adams2017search}
{Adams}, S.~M., {Kochanek}, C.~S., {Gerke}, J.~R., {Stanek}, K.~Z., \& {Dai},
  X. 2017, \mnras, 468, 4968, \dodoi{10.1093/mnras/stx816}

\bibitem[{{Antoni} \& {Quataert}(2023)}]{antoni2023failed}
{Antoni}, A., \& {Quataert}, E. 2023, \mnras, 525, 1229,
  \dodoi{10.1093/mnras/stad2328}

\bibitem[{{Basinger} {et~al.}(2021){Basinger}, {Kochanek}, {Adams}, {Dai}, \&
  {Stanek}}]{basinger2021bh1}
{Basinger}, C.~M., {Kochanek}, C.~S., {Adams}, S.~M., {Dai}, X., \& {Stanek},
  K.~Z. 2021, \mnras, 508, 1156, \dodoi{10.1093/mnras/stab2620}

\bibitem[{Beasor \& Davies(2016)}]{beasor2016evolution}
Beasor, E.~R., \& Davies, B. 2016, Monthly Notices of the Royal Astronomical
  Society, 463, 1269

\bibitem[{Beasor \& Davies(2018)}]{beasor2018evolution}
---. 2018, Monthly Notices of the Royal Astronomical Society, 475, 55

\bibitem[{{Beasor} {et~al.}(2020{\natexlab{a}}){Beasor}, {Davies}, {Smith},
  {van Loon}, {Gehrz}, \& {Figer}}]{beasoe2020new}
{Beasor}, E.~R., {Davies}, B., {Smith}, N., {et~al.} 2020{\natexlab{a}},
  \mnras, 492, 5994, \dodoi{10.1093/mnras/staa255}

\bibitem[{{Beasor} {et~al.}(2020{\natexlab{b}}){Beasor}, {Davies}, {Smith},
  {van Loon}, {Gehrz}, \& {Figer}}]{beasor2020mass}
---. 2020{\natexlab{b}}, \mnras, 492, 5994

\bibitem[{{Beasor} \& {Smith}(2022)}]{beasor2022dersgs}
{Beasor}, E.~R., \& {Smith}, N. 2022, \apj, 933, 41,
  \dodoi{10.3847/1538-4357/ac6dcf}

\bibitem[{Blagorodnova {et~al.}(2017)Blagorodnova, Kotak, Polshaw, Kasliwal,
  Cao, Cody, Doran, Elias-Rosa, Fraser, Fremling,
  {et~al.}}]{blagorodnova2017common}
Blagorodnova, N., Kotak, R., Polshaw, J., {et~al.} 2017, The Astrophysical
  Journal, 834, 107

\bibitem[{{Boersma} {et~al.}(2006){Boersma}, {Hony}, \&
  {Tielens}}]{boersma2006tutau}
{Boersma}, C., {Hony}, S., \& {Tielens}, A.~G.~G.~M. 2006, \aap, 447, 213,
  \dodoi{10.1051/0004-6361:20053904}

\bibitem[{{Bond} {et~al.}(2003){Bond}, {Henden}, {Levay}, {Panagia}, {Sparks},
  {Starrfield}, {Wagner}, {Corradi}, \& {Munari}}]{bond2003v838}
{Bond}, H.~E., {Henden}, A., {Levay}, Z.~G., {et~al.} 2003, \nat, 422, 405,
  \dodoi{10.1038/nature01508}

\bibitem[{{Bradley} {et~al.}(2016){Bradley}, {Sipocz}, {Robitaille},
  {Tollerud}, {Deil}, {Vin{\'\i}cius}, {Barbary}, {G{\"u}nther}, {Bostroem},
  {Droettboom}, {Bray}, {Bratholm}, {Pickering}, {Craig}, {Pascual}, {Greco},
  {Donath}, {Kerzendorf}, {Littlefair}, {Barentsen}, {D'Eugenio}, \&
  {Weaver}}]{bradley2016photutils}
{Bradley}, L., {Sipocz}, B., {Robitaille}, T., {et~al.} 2016, {Photutils:
  Photometry tools}, Astrophysics Source Code Library, record ascl:1609.011.
\newblock \doeprint{1609.011}

\bibitem[{{Chesneau} {et~al.}(2014){Chesneau}, {Meilland}, {Chapellier},
  {Millour}, {van Genderen}, {Naz{\'e}}, {Smith}, {Spang}, {Smoker}, {Dessart},
  {Kanaan}, {Bendjoya}, {Feast}, {Groh}, {Lobel}, {Nardetto}, {Otero},
  {Oudmaijer}, {Tekola}, {Whitelock}, {Arcos}, {Cur{\'e}}, \&
  {Vanzi}}]{ches2014hr}
{Chesneau}, O., {Meilland}, A., {Chapellier}, E., {et~al.} 2014, \aap, 563,
  A71, \dodoi{10.1051/0004-6361/201322421}

\bibitem[{{Davies} \& {Beasor}(2020{\natexlab{a}})}]{davies2020red}
{Davies}, B., \& {Beasor}, E.~R. 2020{\natexlab{a}}, \mnras, 493, 468,
  \dodoi{10.1093/mnras/staa174}

\bibitem[{{Davies} \& {Beasor}(2020{\natexlab{b}})}]{davies2020on}
---. 2020{\natexlab{b}}, \mnras, 496, L142, \dodoi{10.1093/mnrasl/slaa102}

\bibitem[{Davies {et~al.}(2018)Davies, Crowther, \&
  Beasor}]{davies2018humphreys}
Davies, B., Crowther, P.~A., \& Beasor, E.~R. 2018, Monthly Notices of the
  Royal Astronomical Society, 478, 3138

\bibitem[{{Davies} {et~al.}(2022){Davies}, {Plez}, \&
  {Petrault}}]{davies2022superwind}
{Davies}, B., {Plez}, B., \& {Petrault}, M. 2022, \mnras, 517, 1483,
  \dodoi{10.1093/mnras/stac2427}

\bibitem[{Draine \& Lee(1984)}]{draine1984optical}
Draine, B., \& Lee, H.~M. 1984, The Astrophysical Journal, 285, 89

\bibitem[{Draine {et~al.}(2021)Draine, Li, Hensley, Hunt, Sandstrom, \&
  Smith}]{draine2021excitation}
Draine, B., Li, A., Hensley, B.~S., {et~al.} 2021, The Astrophysical Journal,
  917, 3

\bibitem[{F{\"o}rster {et~al.}(2018)F{\"o}rster, Moriya, Maureira, Anderson,
  Blinnikov, Bufano, Cabrera-Vives, Clocchiatti, De~Jaeger, Est{\'e}vez,
  {et~al.}}]{forster2018delay}
F{\"o}rster, F., Moriya, T., Maureira, J., {et~al.} 2018, Nature Astronomy, 2,
  808

\bibitem[{Georgy(2012)}]{georgy2012yellow}
Georgy, C. 2012, Astronomy \& Astrophysics, 538, L8

\bibitem[{{Gerke} {et~al.}(2015){Gerke}, {Kochanek}, \&
  {Stanek}}]{gerke2015search}
{Gerke}, J.~R., {Kochanek}, C.~S., \& {Stanek}, K.~Z. 2015, \mnras, 450, 3289,
  \dodoi{10.1093/mnras/stv776}

\bibitem[{{Greenfield} \& {Miller}(2016)}]{greenfield2016crds}
{Greenfield}, P., \& {Miller}, T. 2016, Astronomy and Computing, 16, 41,
  \dodoi{10.1016/j.ascom.2016.04.001}

\bibitem[{{Guha Niyogi} {et~al.}(2014){Guha Niyogi}, {Min}, {Meixner},
  {Waters}, {Seale}, \& {Tielens}}]{niyogi2014r71}
{Guha Niyogi}, S., {Min}, M., {Meixner}, M., {et~al.} 2014, \aap, 569, A80,
  \dodoi{10.1051/0004-6361/201423746}

\bibitem[{Gustafsson {et~al.}(2008)Gustafsson, Edvardsson, Eriksson,
  J{\o}rgensen, Nordlund, \& Plez}]{gustafsson2008grid}
Gustafsson, B., Edvardsson, B., Eriksson, K., {et~al.} 2008, Astronomy \&
  Astrophysics, 486, 951

\bibitem[{Hosseinzadeh {et~al.}(2023)Hosseinzadeh, Sand, Jencson, Andrews,
  Shivaei, Bostroem, Valenti, Szalai, Burke, Howell, McCully, Newsome,
  Gonzalez, Pellegrino, \& Terreran}]{hosseinzadeh_jwst_2023}
Hosseinzadeh, G., Sand, D.~J., Jencson, J.~E., {et~al.} 2023, ApJL, 942, L18,
  \dodoi{10.3847/2041-8213/aca64e}

\bibitem[{Howitt {et~al.}(2020)Howitt, Stevenson, Vigna-G{\'o}mez, Justham,
  Ivanova, Woods, Neijssel, \& Mandel}]{howitt2020luminous}
Howitt, G., Stevenson, S., Vigna-G{\'o}mez, A., {et~al.} 2020, Monthly Notices
  of the Royal Astronomical Society, 492, 3229

\bibitem[{{Humphreys}(2019)}]{humphreys2019bh1}
{Humphreys}, R.~M. 2019, Research Notes of the American Astronomical Society,
  3, 164, \dodoi{10.3847/2515-5172/ab5191}

\bibitem[{{Jencson} {et~al.}(2023){Jencson}, {Pearson}, {Beasor}, {Lau},
  {Andrews}, {Bostroem}, {Dong}, {Engesser}, {Gomez}, {Guolo}, {Hoang},
  {Hosseinzadeh}, {Jha}, {Karambelkar}, {Kasliwal}, {Lundquist}, {Meza
  Retamal}, {Rest}, {Sand}, {Shahbandeh}, {Shrestha}, {Smith}, {Strader},
  {Valenti}, {Wang}, \& {Zenati}}]{jencson2023ixf}
{Jencson}, J.~E., {Pearson}, J., {Beasor}, E.~R., {et~al.} 2023, \apjl, 952,
  L30, \dodoi{10.3847/2041-8213/ace618}

\bibitem[{{Karambelkar} {et~al.}(2023){Karambelkar}, {Kasliwal},
  {Blagorodnova}, {Sollerman}, {Aloisi}, {Anand}, {Andreoni}, {Brink}, {Bruch},
  {Cook}, {Das}, {De}, {Drake}, {Filippenko}, {Fremling}, {Helou}, {Ho},
  {Jencson}, {Jones}, {Laher}, {Masci}, {Patra}, {Purdum}, {Reedy}, {Sit},
  {Sharma}, {Tzanidakis}, {van der Walt}, {Yao}, \&
  {Zhang}}]{karambelkar2023lrn}
{Karambelkar}, V.~R., {Kasliwal}, M.~M., {Blagorodnova}, N., {et~al.} 2023,
  \apj, 948, 137, \dodoi{10.3847/1538-4357/acc2b9}

\bibitem[{{Kashi} \& {Soker}(2017)}]{kashisoker17}
{Kashi}, A., \& {Soker}, N. 2017, \mnras, 467, 3299,
  \dodoi{10.1093/mnras/stx240}

\bibitem[{Kochanek {et~al.}(2012)Kochanek, Khan, \&
  Dai}]{kochanek2012absorption}
Kochanek, C., Khan, R., \& Dai, X. 2012, The Astrophysical Journal, 759, 20

\bibitem[{{Kochanek}(2015)}]{kochanek2015constraints}
{Kochanek}, C.~S. 2015, \mnras, 446, 1213, \dodoi{10.1093/mnras/stu2056}

\bibitem[{{Kochanek}(2020)}]{kochanek2020on}
---. 2020, \mnras, 493, 4945, \dodoi{10.1093/mnras/staa605}

\bibitem[{Kochanek {et~al.}(2008)Kochanek, Beacom, Kistler, Prieto, Stanek,
  Thompson, \& Y{\"u}ksel}]{kochanek2008survey}
Kochanek, C.~S., Beacom, J.~F., Kistler, M.~D., {et~al.} 2008, The
  Astrophysical Journal, 684, 1336

\bibitem[{{Kochanek} {et~al.}(2023){Kochanek}, {Neustadt}, \&
  {Stanek}}]{kochanek2023bh1}
{Kochanek}, C.~S., {Neustadt}, J.~M.~M., \& {Stanek}, K.~Z. 2023, arXiv
  e-prints, arXiv:2310.01514, \dodoi{10.48550/arXiv.2310.01514}

\bibitem[{{Lovegrove} \& {Woosley}(2013)}]{lovegrove2013very}
{Lovegrove}, E., \& {Woosley}, S.~E. 2013, \apj, 769, 109,
  \dodoi{10.1088/0004-637X/769/2/109}

\bibitem[{MacLeod {et~al.}(2017)MacLeod, Antoni, Murguia-Berthier, Macias, \&
  Ramirez-Ruiz}]{macleod2017common}
MacLeod, M., Antoni, A., Murguia-Berthier, A., Macias, P., \& Ramirez-Ruiz, E.
  2017, The Astrophysical Journal, 838, 56

\bibitem[{{Mason} {et~al.}(2010){Mason}, {Diaz}, {Williams}, {Preston}, \&
  {Bensby}}]{mason2010v1309sco}
{Mason}, E., {Diaz}, M., {Williams}, R.~E., {Preston}, G., \& {Bensby}, T.
  2010, \aap, 516, A108, \dodoi{10.1051/0004-6361/200913610}

\bibitem[{{McDonald} {et~al.}(2022){McDonald}, {Davies}, \&
  {Beasor}}]{mcdonald2022red}
{McDonald}, S. L.~E., {Davies}, B., \& {Beasor}, E.~R. 2022, \mnras, 510, 3132,
  \dodoi{10.1093/mnras/stab3453}

\bibitem[{{Metzger} \& {Pejcha}(2017)}]{metzger2017shock}
{Metzger}, B.~D., \& {Pejcha}, O. 2017, \mnras, 471, 3200,
  \dodoi{10.1093/mnras/stx1768}

\bibitem[{{Munari} {et~al.}(2002){Munari}, {Henden}, {Kiyota}, {Laney},
  {Marang}, {Zwitter}, {Corradi}, {Desidera}, {Marrese}, {Giro}, {Boschi}, \&
  {Schwartz}}]{munari2002v838}
{Munari}, U., {Henden}, A., {Kiyota}, S., {et~al.} 2002, \aap, 389, L51,
  \dodoi{10.1051/0004-6361:20020715}

\bibitem[{{Nadezhin}(1980)}]{nad80}
{Nadezhin}, D.~K. 1980, \apss, 69, 115, \dodoi{10.1007/BF00638971}

\bibitem[{{Neustadt} {et~al.}(2021){Neustadt}, {Kochanek}, {Stanek},
  {Basinger}, {Jayasinghe}, {Garling}, {Adams}, \& {Gerke}}]{neustadt2021lbt}
{Neustadt}, J.~M.~M., {Kochanek}, C.~S., {Stanek}, K.~Z., {et~al.} 2021,
  \mnras, 508, 516, \dodoi{10.1093/mnras/stab2605}

\bibitem[{Ohnaka {et~al.}(2008)Ohnaka, Driebe, Hofmann, Weigelt, \&
  Wittkowski}]{ohnaka2008spatially}
Ohnaka, K., Driebe, T., Hofmann, K.-H., Weigelt, G., \& Wittkowski, M. 2008,
  Astronomy \& Astrophysics, 484, 371

\bibitem[{{Pastorello} {et~al.}(2019){Pastorello}, {Mason}, {Taubenberger},
  {Fraser}, {Cortini}, {Tomasella}, {Botticella}, {Elias-Rosa}, {Kotak},
  {Smartt}, {Benetti}, {Cappellaro}, {Turatto}, {Tartaglia}, {Djorgovski},
  {Drake}, {Berton}, {Briganti}, {Brimacombe}, {Bufano}, {Cai}, {Chen},
  {Christensen}, {Ciabattari}, {Congiu}, {Dimai}, {Inserra}, {Kankare},
  {Magill}, {Maguire}, {Martinelli}, {Morales-Garoffolo}, {Ochner}, {Pignata},
  {Reguitti}, {Sollerman}, {Spiro}, {Terreran}, \&
  {Wright}}]{pastorello2019LRN}
{Pastorello}, A., {Mason}, E., {Taubenberger}, S., {et~al.} 2019, \aap, 630,
  A75, \dodoi{10.1051/0004-6361/201935999}

\bibitem[{{Peeters} {et~al.}(2002){Peeters}, {Hony}, {Van Kerckhoven},
  {Tielens}, {Allamandola}, {Hudgins}, \& {Bauschlicher}}]{peeters2002pah}
{Peeters}, E., {Hony}, S., {Van Kerckhoven}, C., {et~al.} 2002, \aap, 390,
  1089, \dodoi{10.1051/0004-6361:20020773}

\bibitem[{{Pejcha} {et~al.}(2017){Pejcha}, {Metzger}, {Tyles}, \&
  {Tomida}}]{pejcha2017spiral}
{Pejcha}, O., {Metzger}, B.~D., {Tyles}, J.~G., \& {Tomida}, K. 2017, \apj,
  850, 59, \dodoi{10.3847/1538-4357/aa95b9}

\bibitem[{{Perna} {et~al.}(2014){Perna}, {Duffell}, {Cantiello}, \&
  {MacFadyen}}]{perna2014disk}
{Perna}, R., {Duffell}, P., {Cantiello}, M., \& {MacFadyen}, A.~I. 2014, \apj,
  781, 119, \dodoi{10.1088/0004-637X/781/2/119}

\bibitem[{{Prieto} {et~al.}(2009){Prieto}, {Sellgren}, {Thompson}, \&
  {Kochanek}}]{prieto2009300}
{Prieto}, J.~L., {Sellgren}, K., {Thompson}, T.~A., \& {Kochanek}, C.~S. 2009,
  \apj, 705, 1425, \dodoi{10.1088/0004-637X/705/2/1425}

\bibitem[{{Rieke} {et~al.}(2015){Rieke}, {Wright}, {B{\"o}ker}, {Bouwman},
  {Colina}, {Glasse}, {Gordon}, {Greene}, {G{\"u}del}, {Henning}, {Justtanont},
  {Lagage}, {Meixner}, {N{\o}rgaard-Nielsen}, {Ray}, {Ressler}, {van Dishoeck},
  \& {Waelkens}}]{miri2015}
{Rieke}, G.~H., {Wright}, G.~S., {B{\"o}ker}, T., {et~al.} 2015, \pasp, 127,
  584, \dodoi{10.1086/682252}

\bibitem[{{Rieke} {et~al.}(2005){Rieke}, {Kelly}, {Horner}, \& {NIRCam
  Team}}]{nircam}
{Rieke}, M., {Kelly}, D., {Horner}, S., \& {NIRCam Team}. 2005, in American
  Astronomical Society Meeting Abstracts, Vol. 207, American Astronomical
  Society Meeting Abstracts, 115.09

\bibitem[{Scicluna {et~al.}(2015)Scicluna, Siebenmorgen, Wesson, Blommaert,
  Kasper, Voshchinnikov, \& Wolf}]{scicluna2015large}
Scicluna, P., Siebenmorgen, R., Wesson, R., {et~al.} 2015, Online Material p, 1

\bibitem[{{Shara} {et~al.}(2010){Shara}, {Zurek}, {Prialnik}, {Yaron}, \&
  {Kovetz}}]{shara2010v1309sco}
{Shara}, M.~M., {Zurek}, D., {Prialnik}, D., {Yaron}, O., \& {Kovetz}, A. 2010,
  \apj, 725, 824, \dodoi{10.1088/0004-637X/725/1/824}

\bibitem[{Smartt(2015)}]{smartt2015observational}
Smartt, S. 2015, Publications of the Astronomical Society of Australia, 32,
  e016

\bibitem[{Smartt {et~al.}(2009)Smartt, Eldridge, Crockett, \&
  Maund}]{smartt2009death}
Smartt, S., Eldridge, J., Crockett, R., \& Maund, J.~R. 2009, Monthly Notices
  of the Royal Astronomical Society, 395, 1409

\bibitem[{{Smith} \& {Frew}(2011{\natexlab{a}})}]{smith2011etacar}
{Smith}, N., \& {Frew}, D.~J. 2011{\natexlab{a}}, \mnras, 415, 2009,
  \dodoi{10.1111/j.1365-2966.2011.18993.x}

\bibitem[{{Smith} \& {Frew}(2011{\natexlab{b}})}]{sf11}
---. 2011{\natexlab{b}}, \mnras, 415, 2009,
  \dodoi{10.1111/j.1365-2966.2011.18993.x}

\bibitem[{{Smith} \& {Hartigan}(2006)}]{smith2006infra}
{Smith}, N., \& {Hartigan}, P. 2006, \apj, 638, 1045, \dodoi{10.1086/498860}

\bibitem[{Smith {et~al.}(2009)Smith, Hinkle, \& Ryde}]{smith2009red}
Smith, N., Hinkle, K.~H., \& Ryde, N. 2009, The Astronomical Journal, 137, 3558

\bibitem[{Smith {et~al.}(2001)Smith, Humphreys, Davidson, Gehrz, Schuster, \&
  Krautter}]{smith2001asymmetric}
Smith, N., Humphreys, R.~M., Davidson, K., {et~al.} 2001, The Astronomical
  Journal, 121, 1111

\bibitem[{{Smith} {et~al.}(2011{\natexlab{a}}){Smith}, {Li}, {Filippenko}, \&
  {Chornock}}]{smith11}
{Smith}, N., {Li}, W., {Filippenko}, A.~V., \& {Chornock}, R.
  2011{\natexlab{a}}, \mnras, 412, 1522,
  \dodoi{10.1111/j.1365-2966.2011.17229.x}

\bibitem[{{Smith} {et~al.}(2011{\natexlab{b}}){Smith}, {Li}, {Silverman},
  {Ganeshalingam}, \& {Filippenko}}]{smith11lbv}
{Smith}, N., {Li}, W., {Silverman}, J.~M., {Ganeshalingam}, M., \&
  {Filippenko}, A.~V. 2011{\natexlab{b}}, \mnras, 415, 773,
  \dodoi{10.1111/j.1365-2966.2011.18763.x}

\bibitem[{{Smith} {et~al.}(2016){Smith}, {Andrews}, {Van Dyk}, {Mauerhan},
  {Kasliwal}, {Bond}, {Filippenko}, {Clubb}, {Graham}, {Perley}, {Jencson},
  {Bally}, {Ubeda}, \& {Sabbi}}]{smith2016massive}
{Smith}, N., {Andrews}, J.~E., {Van Dyk}, S.~D., {et~al.} 2016, \mnras, 458,
  950, \dodoi{10.1093/mnras/stw219}

\bibitem[{{Smith} {et~al.}(2018){Smith}, {Andrews}, {Rest}, {Bianco}, {Prieto},
  {Matheson}, {James}, {Smith}, {Strampelli}, \& {Zenteno}}]{smith18}
{Smith}, N., {Andrews}, J.~E., {Rest}, A., {et~al.} 2018, \mnras, 480, 1466,
  \dodoi{10.1093/mnras/sty1500}

\bibitem[{{Smolders} {et~al.}(2010){Smolders}, {Acke}, {Verhoelst},
  {Blommaert}, {Decin}, {Hony}, {Sloan}, {Neyskens}, {van Eck}, {Zijlstra}, \&
  {van Winckel}}]{smolders2010when}
{Smolders}, K., {Acke}, B., {Verhoelst}, T., {et~al.} 2010, \aap, 514, L1,
  \dodoi{10.1051/0004-6361/201014254}

\bibitem[{{St{\c{e}}pie{\'n}}(2011)}]{ste2011v1309sco}
{St{\c{e}}pie{\'n}}, K. 2011, \aap, 531, A18,
  \dodoi{10.1051/0004-6361/201116689}

\bibitem[{{Sukhbold} {et~al.}(2018){Sukhbold}, {Woosley}, \&
  {Heger}}]{sukhbold2018high}
{Sukhbold}, T., {Woosley}, S.~E., \& {Heger}, A. 2018, \apj, 860, 93,
  \dodoi{10.3847/1538-4357/aac2da}

\bibitem[{{Tylenda} {et~al.}(2011){Tylenda}, {Hajduk}, {Kami{\'n}ski},
  {Udalski}, {Soszy{\'n}ski}, {Szyma{\'n}ski}, {Kubiak}, {Pietrzy{\'n}ski},
  {Poleski}, {Wyrzykowski}, \& {Ulaczyk}}]{tylenda2011v1309sco}
{Tylenda}, R., {Hajduk}, M., {Kami{\'n}ski}, T., {et~al.} 2011, \aap, 528,
  A114, \dodoi{10.1051/0004-6361/201016221}

\bibitem[{{Umana} {et~al.}(2010){Umana}, {Buemi}, {Trigilio}, {Leto}, \&
  {Hora}}]{umana2010spitz}
{Umana}, G., {Buemi}, C.~S., {Trigilio}, C., {Leto}, P., \& {Hora}, J.~L. 2010,
  \apj, 718, 1036, \dodoi{10.1088/0004-637X/718/2/1036}

\bibitem[{{Van Dyk}(2017)}]{vandyk2017direct}
{Van Dyk}, S.~D. 2017, Philosophical Transactions of the Royal Society of
  London Series A, 375, 20160277, \dodoi{10.1098/rsta.2016.0277}

\bibitem[{Xie \& Ho(2022)}]{xie2022ionization}
Xie, Y., \& Ho, L.~C. 2022, The Astrophysical Journal, 925, 218

\end{thebibliography}
\bibliographystyle{aasjournal}



\end{document}